\documentclass[useAMS,usenatbib,usegraphicx]{mn2e}
\usepackage{aas_macros}
\usepackage{amssymb}

\title[The mass of dark matter haloes]{The virialized mass of dark matter haloes}
\author[A.~J. Cuesta et al.]{A.~J. Cuesta,$^1$\thanks{E-mail: ajcv@iaa.es} F. Prada,$^1$ A. Klypin$^2$ and M. Moles$^1$\\
$^1$Instituto de Astrof{\'\i}sica de Andaluc{\'\i}a (CSIC),
Camino Bajo de Hu\'etor 50, E-18008 Granada, Spain \\
$^2$Department of Astronomy,
New Mexico State University, Las Cruces, NM 88001, USA
}

\begin{document}
\bibliographystyle{mn2e}
\maketitle

\begin{abstract}
Virial mass is used as an estimator for the mass of a dark matter halo. However, the commonly used constant overdensity criterion does not reflect the dynamical structure of haloes. Here we analyze dark matter cosmological simulations in order to obtain properties of haloes of different masses focusing on the size of the region with zero mean radial velocity. Dark matter inside this region is stationary, and thus the mass of this region is a much better approximation for the virial mass. We call this mass the static mass to distinguish from the commonly used constant overdensity mass. We also study the relation of this static mass with the traditional virial mass, and we find that the matter inside galaxy-size haloes ($M\approx 10^{12}{\rm M}_{\sun}$) is underestimated by the virial mass by nearly a factor of two. At $z\approx 0$ the virial mass is close to the static mass for cluster-size haloes ($M\approx 10^{14}{\rm M}_{\sun}$). The same pattern -- large haloes having $M_{\rm vir} > M_{\rm static}$ -- exists at all redshifts, but the transition mass $M_0 = M_{\rm vir} = M_{\rm static}$ decreases dramatically with increasing redshift: $M_0(z) \approx 3\times 10^{15}h^{-1}{\rm M}_{\sun} (1+z)^{-8.9}$. When rescaled to the same $M_0$ haloes clearly demonstrate a self-similar behaviour, which in a statistical sense gives a relation between the static and virial mass. To our surprise we find that the abundance of haloes with a given static mass, i.e. the static mass function, is very accurately fitted by the Press \& Schechter approximation at $z=0$, but this approximation breaks at higher redshifts $z\simeq 1$. Instead, the virial mass function is well fitted as usual by the Sheth \& Tormen approximation even at $z\lesssim 2$. We find an explanation why the static radius can be 2--3 times larger as compared with the constant overdensity estimate. The traditional estimate is based on the top-hat model, which assumes a constant density and no rms velocities for the matter before it collapses into a halo. Those assumptions fail for small haloes, which find themselves in environment, where density is falling off well outside the virial radius and random velocities grow due to other haloes. Applying the non-stationary Jeans equation we find that the role of the pressure gradients is significantly larger for small haloes. At some moment it gets too large and stops the accretion.
\end{abstract}

\begin{keywords}
dark matter -- galaxies: haloes -- large-scale structure of Universe -- cosmology: theory -- methods: $N$-body simulations
\end{keywords}

\section{Introduction}
The currently accepted paradigm of hierarchical clustering (\citealt{Whi78}, \citealt{Blu84}) provides a picture of assembly of dark matter haloes in which more massive haloes are formed through merging and accretion of smaller ones. This picture is supported for example by recent observations of ongoing mergers in clusters (e.g. \citealt{vDok99}, \citealt{Rin07}). The theoretical framework of the cold dark matter (CDM) has been proved successful in the description of the structure formation in the Universe from tiny fluctuations in the primordial density field (see \citealt{Pri03} for a review). This model has received support from many different observations, in particular of the gravitational lensing effect (\citealt{Smi01}, \citealt{Guz02}, \citealt{Kne03}, \citealt{Hoe04}, \citealt{She04}, \citealt{Man06}), CMB \citep{Spe07}, the abundance of clusters (\citealt{Pie01}, \citealt{Gla07}), and satellite dynamics (\citealt{Zar94}, \citealt{Pra03}). Cosmological simulations with ever increasing resolution play important role by making accurate predictions of different properties of dark matter haloes. Results of those simulations are used by other methods. For example, semi-analytical models of galaxy formation have been either incorporated into N-body simulations or use statistics such as halo mass function and merging trees, which were calibrated and tested using the simulations (e.g. \citealt{Som99,Cro06}). The simulations reveal important information about the internal structure of dark matter haloes (e.g. \citealt{Nav97}, \citealt{Bul01b}, \citealt{Tay01}) which is reflecting their underlying dynamics.

Many results of simulations implicitly use some definition of what is the size and mass of a collapsed dark matter halo. The problem is that there is no well-defined boundary of a halo: density field is smooth around the halo. The common prescription for this boundary (and hence the mass belonging to the halo) is defined through the spherical collapse model (\citealt{Gun72}, \citealt{Gun77}). The size of the halo at redshift $z$ is given by the half of the radius of the spherical shell at turnaround which is collapsing at that redshift. This is the \textit{virial radius}. Thus, it is very common to measure the mass of haloes in cosmological simulations taking the particles inside a sphere of fixed spherical overdensity or to take those particles which are connected by a inter-particle separation below a given value (the friends-of-friends algorithm). We note that there is little justification for using the top-hat collapse model. Haloes do not collapse from perfect spherical homogeneous distribution. The environment of haloes is typically very non-spherical with most of accretion happening from few elongated filaments. The random velocities of the accreted matter also cannot be neglected. As the fluctuations collapse, the dark matter, which is being accreted, increases it rms velocities. This effective pressure should affect the accretion rate. The only motivation for using the top-hat model comes from simulations. Indeed, early simulations indicated that the radius of overdensity 178 is close to the virial radius \citep{Col96}: "the radius $r_{178}$ approximately demarcates the inner regions of haloes at $r\lesssim r_{178}$ which are in approximate dynamical equilibrium from the outer regions at $r\gtrsim r_{178}$ which are still infalling". Thus, the radius of overdensity 200 ($\approx 178$) became the virial radius. The reason why this was a good approximation is simply coincidental: the early simulations were mostly done for cluster-size haloes and, indeed, for those masses the virial radius is close to the radius of overdensity 200. The early models were flat models without the cosmological constant. Models with the cosmological constant have produced significant confusion in the community. The top-hat model must be modified to incorporate the changes due to the different rate of expansion and due to the different rate of growth of perturbations (see \citealt{Pri97}). That path produced the so called virial radius, which for the standard cosmological model gives the radius of overdensity relative to matter of about 340 \citep{Bul01}. Still, a large group of cosmologists uses the old overdensity 200 relative to the critical density even for the models with the cosmological constant.

In this paper, we cast some light on this subject by searching for a physical extent of dark matter haloes, which is related to the physical processes that occur around collapsed structures. \citet{Pra06} provided first results, which indicated that spherically averaged, mean radial velocity profiles show an inner region in which there is no net infall or outflow. The size of this region in virial units is mass-dependent: for galactic-size haloes it may even reach three times the virial radius. We use this result to study the properties of the mass inside this region and in particular, to determine the redshift evolution of such mass. Recent effort (\citealt{Wec02}, \citealt{Zha03}) has been devoted to the analysis of the evolution of virial mass inside dark matter haloes, sometimes referred to as the mass accretion history. However, the picture presented in these works has been recently put into question. The analysis of the mass inside a fixed physical radius has revealed that galaxy-size haloes experience unphysical growth of their virial mass \citep{Die07}. The mean background density decreases as the Universe expands, making the virial radius to increase even in the case of no accretion or mergers. This is clearly an artifact of the definition. Other works realized this issue, for example, the evolution of the spin parameter when the matter inside a fixed radius is taken into account differs from that when using the evolving $R_{\rm vir}$ instead \citep{Don07}. On the other hand, there is another effect apart from accretion and merging, which has not been taken into account: haloes may also grow via relaxation of the surrounding regions near them. This is an important effect as it is a reflection of the dynamical processes which are turning non-virialized mass in the outskirts of a halo into the mass associated to it. Moreover, the long-term evolution of dark matter haloes show an interesting feature for $\Lambda$CDM cosmologies: their mass turns out to converge to an asymptotic value which depends on the definition of mass, as pointed out by \citet{Bus05}. Thus, it is desirable that the mass of a halo is measured using a virialization-based criteria (see \citealt{Mac03} for an interesting approach), instead of using boundaries of a given overdensity.

The interest of the measurement of the physical mass associated to dark matter haloes is not only theoretical. Indeed, it may have a great impact on the number of collapsed objects in a given range of mass, i.e. the mass function \citep{Whi01}. Besides, it is of great importance for the process of formation and evolution of galaxies, and, hence, it is relevant for the results obtained from semi-analytical modelling of galaxy formation (e.g. \citealt{Cro06}), to predict the main properties of observed galaxies. Thus, it turns out to be mandatory to bring attention on the mass belonging to a halo, if accurate predictions of the physics behind galaxies from cosmological simulations are to be drawn.

This paper is organized as follows: in Section~\ref{sec:simul} we describe the set of cosmological simulations used in our analysis and the properties of the halo samples. In Section~\ref{sec:static} we present the definition of static mass and how it is related to objects of different sizes. A brief analysis of equilibrium in this context is presented in Section~\ref{sec:jeans}. The scaling properties of the static to virial mass relation with redshift are shown in Section~\ref{sec:scaling}, together with a simple model derived from this scaling relation. We obtain the static mass function and compare it with analytical models in Section~\ref{sec:mf}. In Section~\ref{sec:evolution} we present the evolution of the static mass tracking the halo progenitors. We discuss our results in Section~\ref{sec:discussion} and present our conclusions in Section~\ref{sec:conclusion}.

\section{$N$-body simulations and halo selection}
\label{sec:simul}
We use four different high-resolution $\Lambda$CDM simulations, which were run using the Adaptive Refinement Tree (ART) code \citep{Kra97}. These cosmological simulations are selected in order to span a broad range of halo masses with very good statistics of the number of haloes. The values of the cosmological parameters are $\Omega_0=0.3$, $\Omega_{\rm bar}=0.045$, $\Omega_{\Lambda}=0.7$, $\sigma_8=0.9$ and $h=0.7$. Main simulation parameters may be found in Table~\ref{tab:boxes}.

\begin{table*}
\caption{The main parameters of the cosmological simulations.}
\begin{tabular}{l|cccc}
\hline
        & Box80S & Box80G & Box120 & Box250c \\
\hline
Box Size ($h^{-1}Mpc$) & 80 & 80 & 120 & 250 \\
Mass of a particle ($h^{-1}{\rm M}_{\sun}$) & $4.91\times 10^6$ & $3.18\times 10^8$ & $1.07\times 10^9$ & $9.67\times 10^9$ \\
Spatial Resolution ($h^{-1}$kpc) & 0.52 & 1.2 & 1.8 & 7.6 \\
Number of particles & 159.8M & $512^3$ & $512^3$ & $512^3$ \\
\hline
\end{tabular}
\label{tab:boxes}
\end{table*}

The resolution and the halo mass range covered by these simulations allow us to perform a detailed analysis of the properties of dark matter haloes with very good statistics. Dark matter haloes are identified using the Bound Density Maxima halo finder (BDM, \citealt{Kly99}). In order to resolve the inner structure and dynamics of dark matter haloes, only haloes which enclose more than $2,000$ particles inside the virial radius at $z=0$, were chosen. We take into account all the mass (both bound and unbound particles) inside the virial radius of the halo. This radius is defined here as the radius of the sphere enclosing an overdensity given by \citet{Bry98}, i.e.:
\begin{equation}
\Delta_c=18\pi^2+82x-39x^2 \quad x\equiv\Omega(z)-1
\end{equation}

\begin{figure}
\includegraphics[width=0.5\textwidth]{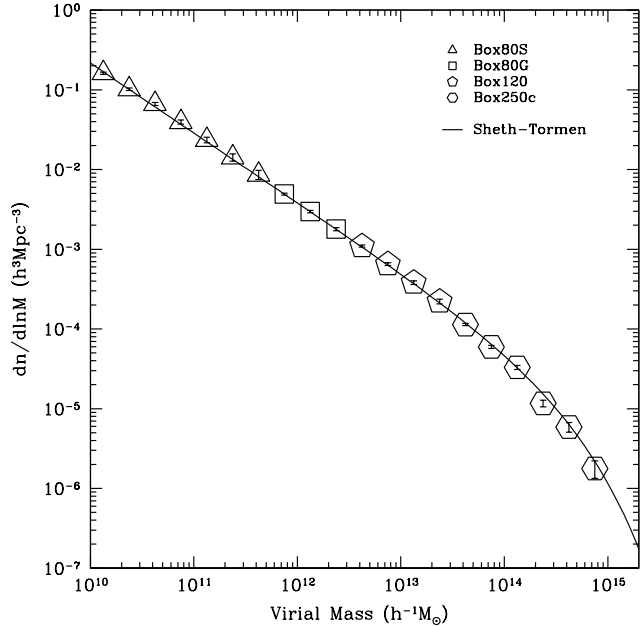}
\caption{The virial mass function using cosmological simulations listed in Table \ref{tab:boxes}. The error bars inside the symbols represent the Poissonian error. The analytical mass function of Sheth and Tormen provides a very good fit for our data.}
\label{fig:mf}
\end{figure}

In Figure~\ref{fig:mf} we plot the virial mass function of dark matter haloes over five orders of magnitude. Here the error bars represent the Poissonian error, showing very good statistics for all except for the highest mass bins. The different symbols shows which simulation is used for each mass bin. The relation between mass bins and simulations depends on the fact that we only choose haloes enclosing more than $2,000$ particles inside the virial radius at $z=0$. If two different simulations pass this constraint, then we choose the one with highest statistics. As previously shown (e.g. \citealt{She01}), the virial mass function in cosmological simulations is well fitted by the Sheth and Tormen function if $a\simeq 0.707$ (\citealt{She99}, hereafter ST) in the mass range under study. These four simulations, as mentioned above, cover a broad range in mass, from $10^{10}h^{-1}{\rm M}_{\sun}$ low mass haloes to cluster-size haloes with nearly $10^{15}h^{-1}{\rm M}_{\sun}$.

In order to select a sample of haloes for our analysis, we take all the haloes whose centres are not inside the virial radius of a larger halo (hereafter \textit{distinct} haloes). This criterion reduces strong environmental effects: internal dynamics of haloes is less affected by massive neighbours.

\section{Static mass of dark matter haloes}
\label{sec:static}

\begin{figure}
\includegraphics[width=0.5\textwidth]{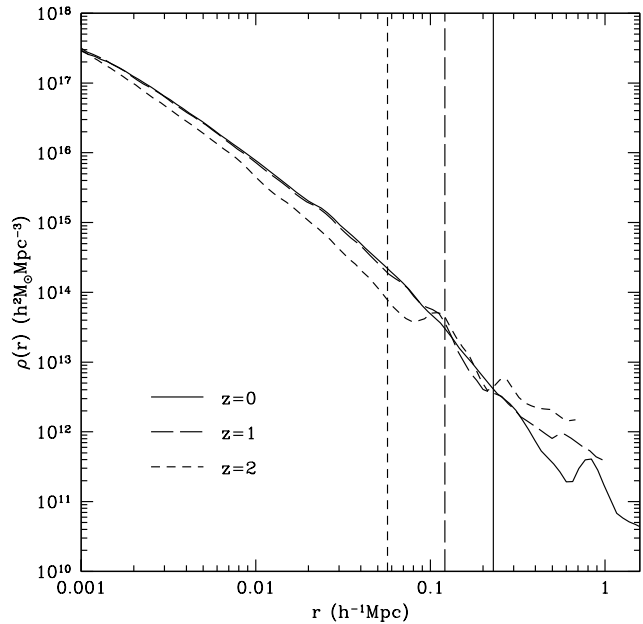}
\caption{Evolution of the density profile for a galactic-size halo. The vertical lines mark the virial radius at each redshift. The change in the virial radius by a factor of two since redshift $z=1$ gives a false impression of a strong evolution of the halo. Yet, there was a very little evolution in the physical density in this halo since $z=1$.}
\label{fig:dens}
\end{figure}

In this Section we introduce and motivate our definition of static mass. When using cosmological simulations, it is common to define the mass belonging to dark matter haloes in terms of a given spherical overdensity or inter-particle distance. However, so far, it is computationally challenging to define a real virial mass, i.e. the mass which is showing features of statistical equilibrium, for a large number of haloes. This has to be kept in mind, if accurate results are to be found, although there is some kind of agreement between different definitions based on overdensity \citep{Whi01}. A particular case in which an overdensity-based definition may fail in providing a faithful description of dark matter haloes is the mass accretion history. Figure~\ref{fig:dens} clearly demonstrates this. Here we show the evolution of the density profile of a very high resolution (more than two million particles inside the virial radius) galaxy-size halo from our cosmological simulation Box20, which has a better spatial (0.152 $h^{-1}$kpc) and mass resolution ($6.14\times 10^{5}$ $h^{-1}{\rm M}_{\sun}$, using 8.98M particles) with the same cosmological parameters as the simulations mentioned above. It is clear that the density profile does not evolve in a significant way from $z=1$ to $z=0$, as already pointed out by \citet{Die07}. Nonetheless, the virial radius varies by about a factor of two as marked with vertical lines in Figure~\ref{fig:dens}. This growth of the virial radius is obviously an artifact of overdensity-based mass that does not reflect the real accretion of mass in the halo but the evolution of the mean matter density of the Universe, $\rho_m(z)=(1+z)^3\rho_{m,0}$.

\begin{figure}
\begin{center}
\includegraphics[width=0.39\textwidth]{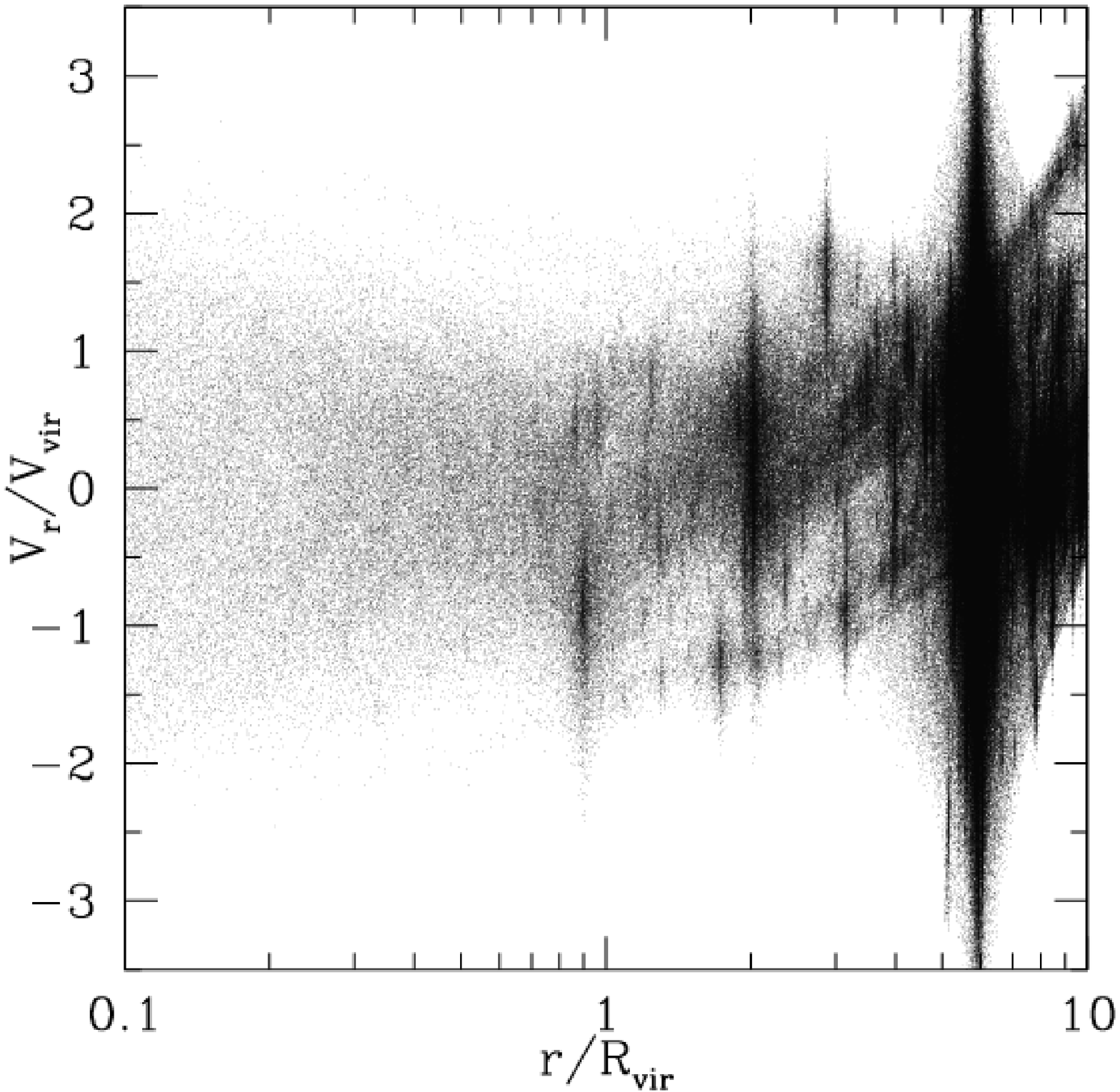} \\
\includegraphics[width=0.39\textwidth]{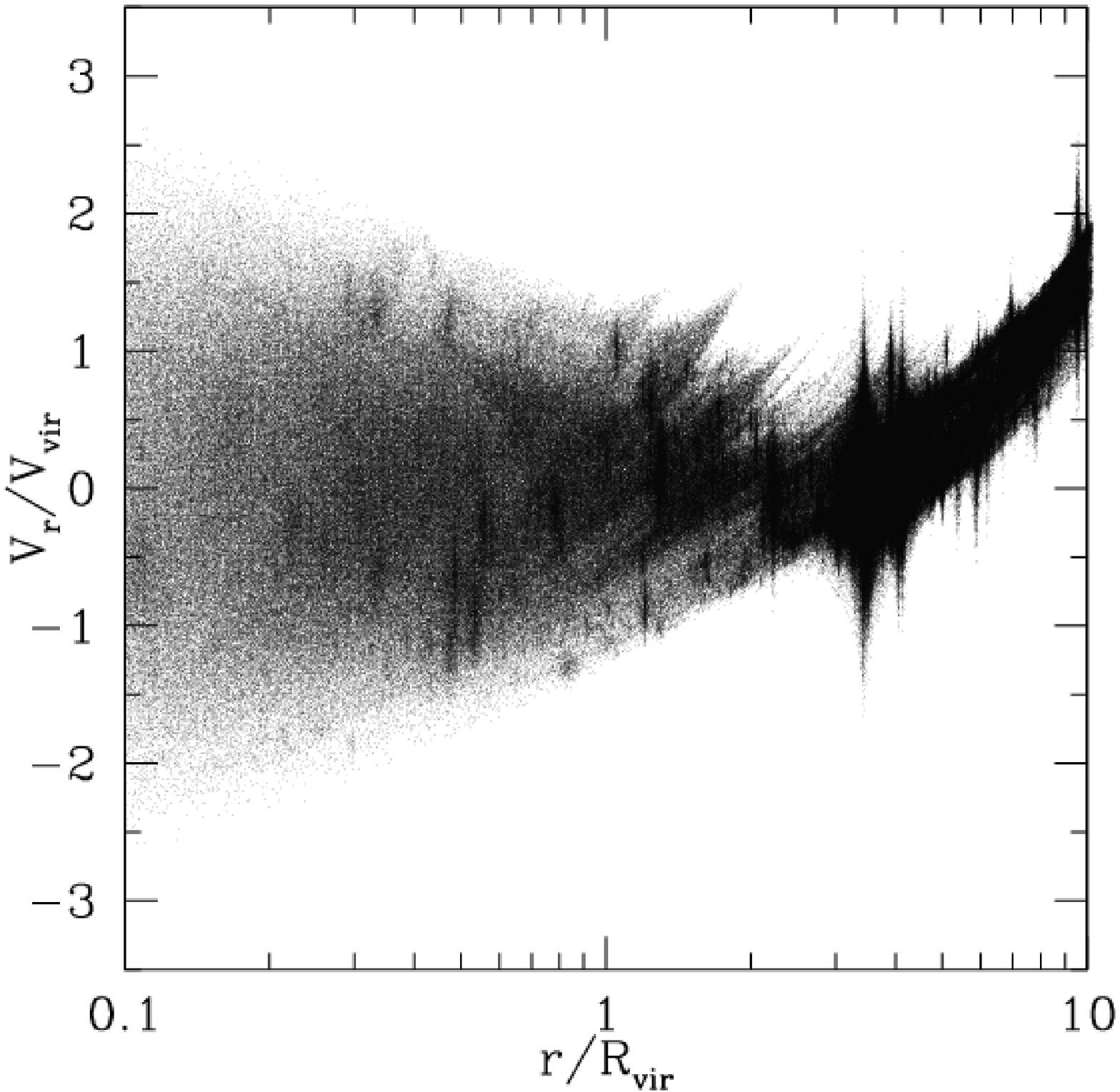} \\
\includegraphics[width=0.39\textwidth]{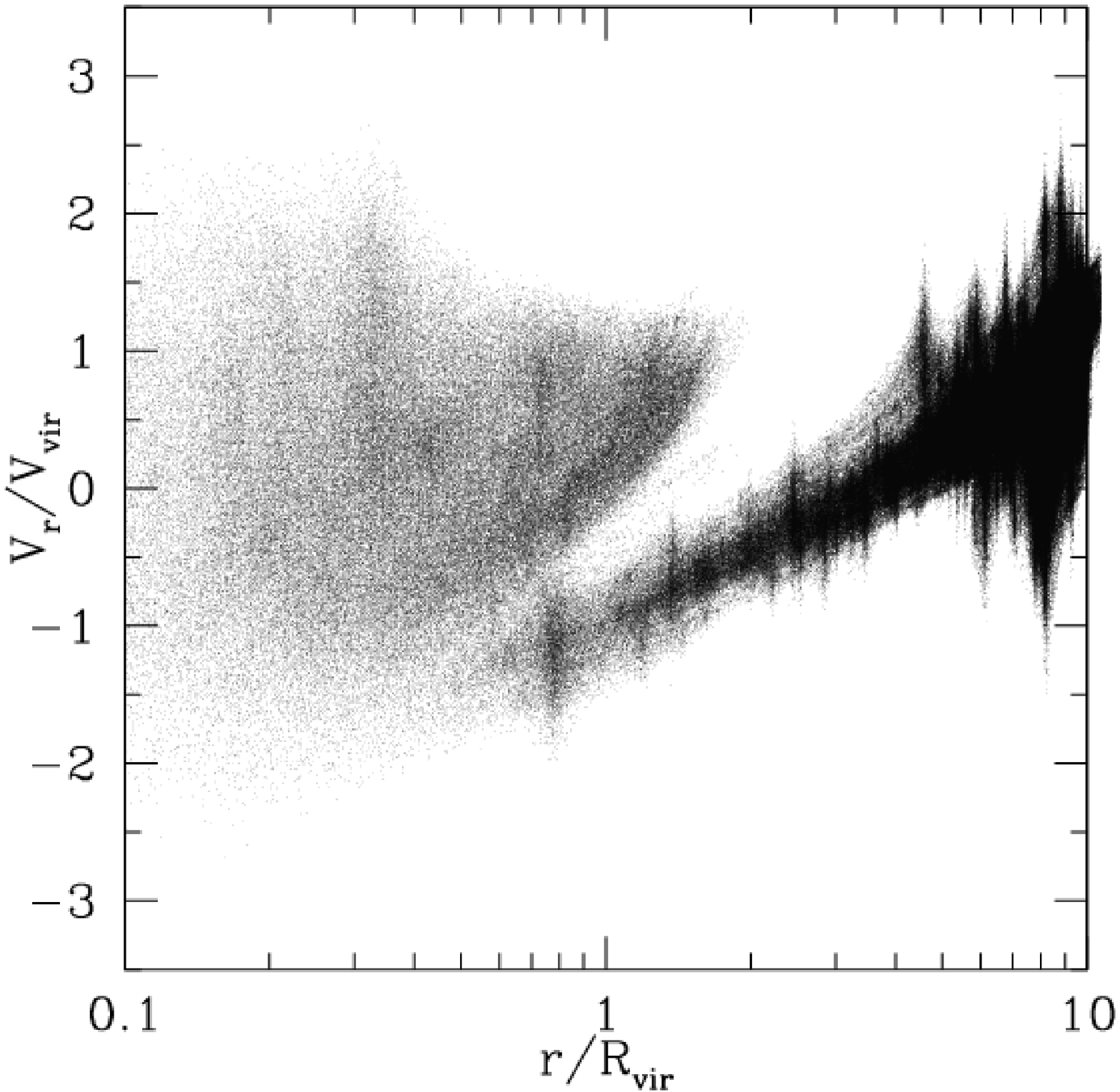} \\
\end{center}
\caption{Phase-space diagram for the particles in dark matter haloes. The top panel represent a low-mass halo ($M_{\rm vir}=2.9\times 10^{11}h^{-1}{\rm M}_{\sun}$) and the central panel displays a galactic-size halo ($M_{\rm vir}=1.4\times 10^{12}h^{-1}{\rm M}_{\sun}$), whereas the bottom panel shows a cluster-size halo ($M_{\rm vir}=1.3\times 10^{15}h^{-1}{\rm M}_{\sun}$). Clusters clearly show strong infall pattern around the virial radius, whereas this is not observed in smaller size haloes.}
\label{fig:cloud}
\end{figure}

Therefore, a detailed analysis of the structure of dark matter haloes may prove to be valuable in the search for the virialized regions of the haloes. For example, the study of the radial phase-space diagram provides some insights on halo structure. In Figure~\ref{fig:cloud} we show the radial velocity pattern of dark matter particles inside haloes of different masses taken from our simulation boxes. The top, middle, and bottom panels are for a low-mass halo with $M_{\rm vir}=2.9\times 10^{11}h^{-1}{\rm M}_{\sun}$, a galactic-size halo of mass $M_{\rm vir}=1.4\times 10^{12}h^{-1}{\rm M}_{\sun}$ and a cluster-size halo with $M_{\rm vir}=1.3\times 10^{15}h^{-1}{\rm M}_{\sun}$, respectively. About $20$ per cent randomly selected particles are shown for the low-mass halo and the galactic-size halo, and $100$ per cent for the cluster-size halo. The structure of haloes in phase-space was already discussed in \citet{Bus05}. In the inner parts the average radial velocities are zero. Around the virial radius there are some signs of infall or outflow, as it will be discussed below. At large distances, the Hubble flow is the most remarkable feature, except for some spikes showing the presence of neighbouring haloes. This plot shows a substantial difference in the structure of dark matter haloes associated with different masses. Although the transition from the inner region to the Hubble flow shows monotonically increasing radial velocity in the low-mass and the galactic-size halo, the situation is different for clusters where there is a significant tail in the velocity distribution with large negative velocities (see also Fig.~3 in \citealt{Bus03}). This infall is altering the equilibrium of the particle distribution in the halo, and hence that region is no longer virialized, as already pointed out by \citet{Mac03}. This suggests a natural choice for the boundary of a dark matter halo, which is defined by the innermost radius in which mean radial velocity is equal to zero. In practice, we use a small threshold in order to reduce effects of the statistical noise. The deviation of the spherically averaged radial velocity profile from zero is a signature of non-virialization. For this reason the zero-velocity radius is defining the mass, which is associated to a halo. This radius will be hereafter referred to as the static radius, and the mass inside a sphere of this radius, as the static mass of the halo.

Although the threshold in the mean radial velocity is somewhat arbitrary and may affect our results, we take advantage of the good statistics in our simulation boxes to reduce this threshold to only 5 per cent of the virial velocity $V_{\rm vir}\equiv\sqrt{GM_{\rm vir}/R_{\rm vir}}$ of the halo. This allows us to get acceptable signal-to-noise ratios in the average profiles for all the mass ranges under study, without considerable overestimation of the virialized region.

Individual haloes usually have similar radial profiles at a given range of mass. Thus, the analysis of the average halo profiles is reasonable. The radial profile of a given physical quantity is then obtained by getting the median value of this quantity over all the haloes in the sample for a given mass bin. This is done for every radial bin, which are logarithmically separated for convenience. For the innermost radial bins, there is an additional constraint: the median profile is representative of the halo sample at a radial distance $R$ only if a large fraction (about 80 per cent) of the haloes in this sample contain at least 200 particles \citep{Kly01} inside the sphere of radius $R$.

\begin{figure}
\includegraphics[width=0.5\textwidth]{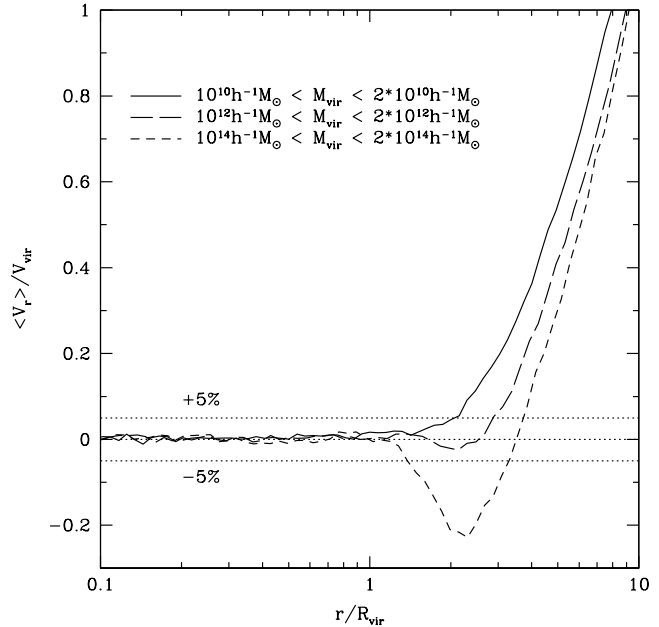}
\caption{Mean radial velocity for three different mass bins. The profiles were obtained by averaging over hundreds of distinct haloes on each mass bin. In dotted line is shown the selected threshold delimiting the static region (5 per cent of the virial velocity). Cluster-size haloes display a region with strong infall (dashed line). On the contrary, low-mass haloes (solid line) and galactic haloes (long-dashed line) do not show infall at all but a small outflow preceding the Hubble flow.}
\label{fig:vr}
\end{figure}

As mentioned above, there is an important mass dependence in the radial velocity distribution of the haloes. The analysis of the median radial velocity profile reveals a prominent infall for cluster-size haloes and a noticeable outflow for low-mass haloes, whereas galaxy-size haloes show a balanced situation between infall and outflow (see \citealt{Pra06} for more details). This is represented in Figure~\ref{fig:vr}. This general trend is dividing the total sample in two sets of haloes: those which are exceeding the threshold from below (infall), and those which exceed it from above (outflow).

\begin{figure}
\includegraphics[width=0.5\textwidth]{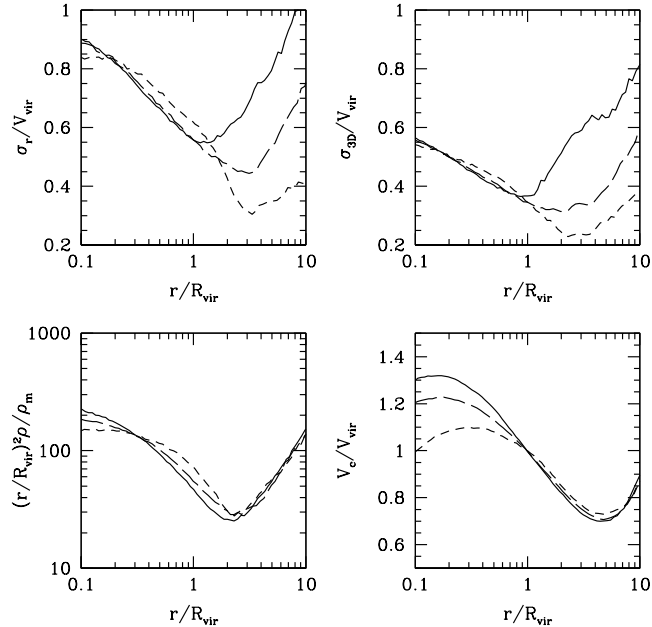}
\caption{Median profiles for the same halo mass bins as in Figure~\ref{fig:vr}. These profiles show how the behaviour of haloes depends on the halo mass. Top left panel: radial velocity dispersion. Top right: 3D velocity dispersion. Bottom left: density profile. Bottom right: Circular velocity profile. The different line styles represent the same mass bins as in Figure~\ref{fig:vr}.}
\label{fig:profiles}
\end{figure}

Other properties of haloes also depend on halo mass. Figure~\ref{fig:profiles} shows some of them for the same mass bins as in Figure~\ref{fig:vr}. Upper panels represent the radial and 3D velocity dispersions (see also \citealt{Woj05} for an extensive analysis of radial velocity moments in massive haloes, and \citealt{Woj08} for a phenomenological model which describes these). The radial velocity dispersion for haloes with higher concentration has a larger inner value in terms of the virial velocity \citep{Lok01}. In the outer region there is a local minimum, which radius and magnitude also show dependence on halo mass. In the bottom left panel we show the median density profiles multiplied by $\left(r/R_{\rm vir}\right)^2$ in order to make more clear any change in their logarithmic slope. At smaller distances $r<0.1R_{\rm vir}$ (not shown in the figure) curves start to decline because the slope is getting smaller than $-2$. It is also noticeable that beyond 1--2 $R_{\rm vir}$ the density profiles do not decrease as $r^{-3}$ as for the NFW density profile. For a detailed parametrization far beyond $R_{\rm vir}$ see Tav{\'\i}o et al. (in preparation). Finally, the bottom right panel shows the radial profile of the circular velocity, which by definition coincides with the virial velocity at $r=R_{\rm vir}$. The maximum of this profile is again mass-dependent, as it is somewhat related to the inner concentration.

\begin{figure}
\includegraphics[width=0.5\textwidth]{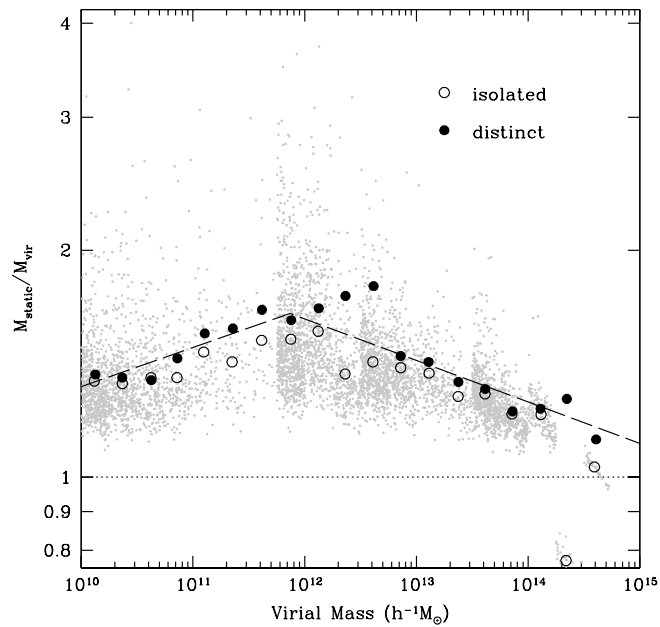}
\caption{The halo static mass -- virial mass relation at $z=0$. Dots show the relation for individual isolated haloes. Circles show this relation for isolated (open) and distinct (solid) median haloes at different mass bins. The ratio $M_{\rm static}/M_{\rm vir}$ increases due to reduced outflow up to the first appearance of infall, making this ratio to decrease. Dashed line shows the approximation to two power laws connected where they cross each other, the parameters are those listed in Table~\ref{tab:fit}.}
\label{fig:ms}
\end{figure}

The analysis of the mean radial velocity profiles over all masses (similar to those in Figure~\ref{fig:vr}) allows us to draw the relation between the static mass and the virial mass. This relation is shown in Figure~\ref{fig:ms}. Solid circles represent the $M_{\rm static}/M_{\rm vir}$ ratio for the median profile built from all distinct haloes, while open circles show the same ratio for the median profile of isolated haloes. The isolation criteria here is that the nearest neighbour of a halo is further than three times the sum of the virial radii of both haloes. The uncertainty on this relation is difficult to estimate, as static masses could not be obtained for individual haloes. Our radial velocity profiles are too noisy (because of infalling and outflowing subhaloes) to distinguish the static region, and hence we stacked these profiles in order to analyse them. Thus, we calculate the halo-to-halo variation of this relation by measuring for each halo the mass inside the median static radius. In this way we obtain the $M_{\rm static}/M_{\rm vir}$ ratio for isolated haloes, shown in this figure as dots. The distribution of this ratio has a long tail to large values of this ratio due to surrounding neighbours, which are inside the median static radius. Again, these unrealistic values of $M_{\rm static}/M_{\rm vir}$ are due to the lack of a reliable determination of the static mass for individual haloes. There are no large differences between distinct and isolated haloes (except for those bins with poor statistics), so we take the $M_{\rm static}/M_{\rm vir}$ relation for distinct haloes, as the isolation criteria reduces their number drastically.

The dependence of this relation with the virial mass is straightforward. The outflow in low-mass haloes is less prominent with
increasing mass, and hence the ratio $M_{\rm static}/M_{\rm vir}$ increases. A drop in this ratio occurs due to the transition to infall around $5\times10^{12}h^{-1}{\rm M}_{\sun}$, and then it decreases due to higher infall with increasing virial mass. As it may be readily checked in Figure~\ref{fig:vr}, the lowest mass in which infall takes place is very threshold-dependent. To alleviate this, we assume that this relation is a continuous function which is fairly well approximated by two power laws:

\begin{equation}
\log_{10}(M_{\rm static}/M_{\rm vir})=
\left\lbrace\begin{array}{lr}
\alpha_1\log_{10}(M_{\rm vir})+\beta_1 & \mathrm{region\; 1} \\
\alpha_2\log_{10}(M_{\rm vir})+\beta_2 & \mathrm{region\; 2}\\
\end{array}\right.
\end{equation}
where the mass is measured in units of $h^{-1}{\rm M}_{\sun}$. In Table~\ref{tab:fit} we list the values of the best-fitting parameters for $\alpha$ and $\beta$.

\begin{table*}
\caption{Parameters of the approximation of the $M_{\rm static}/M_{\rm vir}$ ratio as a function of the virial mass.}
\begin{tabular}{lccc}
\hline
region & mass range & $\alpha$ & $\beta$ \\
\hline
1 & $1.0\times10^{10} \leq M_{\rm vir}/h^{-1}{\rm M}_{\sun}\leq 7.4\times10^{11}$ &$\alpha_1=+0.052\pm0.005$ & $\beta_1= -0.40\pm0.05$ \\
2 & $7.4\times10^{11}\leq M_{\rm vir}/h^{-1}{\rm M}_{\sun}\leq 5.0\times10^{14}$ &$\alpha_2=-0.055\pm0.009$ & $\beta_2= +0.87\pm0.12$ \\
\hline
\end{tabular}
\label{tab:fit}
\end{table*}

Interestingly, concentration of individual dark matter haloes decreases with static mass, in a similar way as it does with $M_{\rm vir}$. This makes the ratio $M_{\rm static}/M_{\rm vir}$ non sensitive to halo concentration. Again, the static mass of individual haloes is estimated by measuring the mass inside their median static radius corresponding to its virial mass bin as in Figure~\ref{fig:ms}. The $M_{\rm static}(c)$ dependence is similar to $M_{\rm vir}(c)$ because $M_{\rm static}$ is close to $M_{\rm vir}$ (the difference is less than a factor of two in the mass range under study). Hence correlations involving the logarithm of the mass can only change their zero point by $\sim 0.3$ if we use the static mass instead of the virial mass. The values of the parameter $\alpha$ in Table~\ref{tab:fit} suggest that the slope would present also a small change of $\sim 5\%$ which might be important in some correlations. In the particular case of the correlation between halo mass and halo concentration (e.g. \citealt{Mac07}), the replacement of virial mass to static mass introduces a modification which is small enough so that this correlation remains almost entirely unaffected.

\section{Analysis of Equilibrium}
\label{sec:jeans}

\begin{figure}
\includegraphics[width=0.5\textwidth]{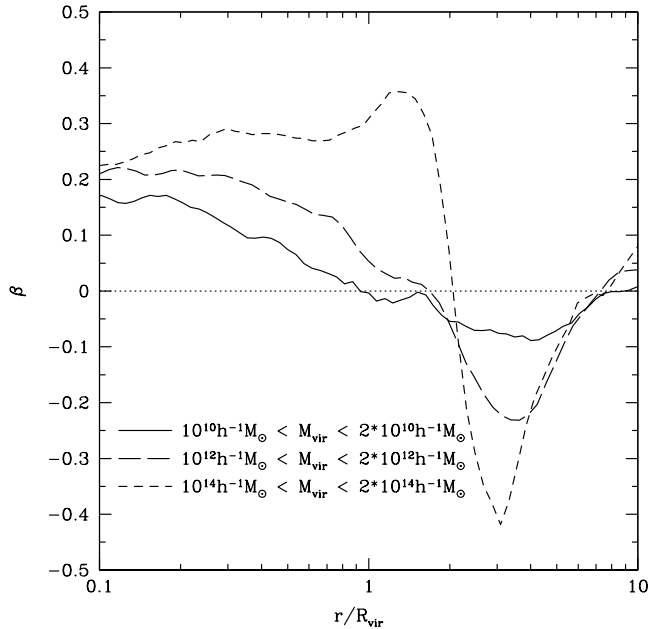}
\caption{Velocity anisotropy profile for the same mass bins as in Figure~\ref{fig:vr} and Figure~\ref{fig:profiles}. In the inner region the velocity anisotropy is preferentially radial. It changes to nearly isotropic in the outer regions with $R\approx (1-2)R_{\rm vir}$ followed by preferentially tangential at larger distances.}
\label{fig:beta}
\end{figure}

Haloes of different mass clearly show very different structure. At small ($r< R_{\rm vir}$) distances the main differences are due to the differences in concentration. The haloes inside the virial radius are nearly in equilibrium (with possible exception of the most massive clusters). At larger distances the differences are more complicated. Figure~\ref{fig:vr} and similar results in \citet{Pra06} demonstrate that haloes of small mass have static radius few times larger than the formal virial radius. To make things even more complicated, the small haloes do not have an infall region. We can try to somewhat clarify the situation by applying the Jeans equation. In order to do this, we first need to measure the velocity anisotropy
$\beta=1-\sigma_t^2/\left(2\sigma_r^2\right)$.

In Figure~\ref{fig:beta} we show the median radial profile for $\beta$ for the same mass bins as in Figure~\ref{fig:vr}. Inside the virial radius $\beta$ is positive. Cluster-mass haloes have a familiar behaviour: $\beta$ increases with distance and gets to $\beta \approx 0.3-0.4$ around the virial radius. The real surprise came for the small galaxy-size haloes. After initial increase, $\beta$ reaches maximum $\beta\approx 0.2$ and then declines with distance. It goes almost to zero at $r = (1-2)R_{\rm vir}$. At larger distances the velocity anisotropy is negative and approaches zero at very large distances. We should note that $\beta=0$ does not mean nearly circular orbits. Orbits of dark matter particles or satellites are typically very elongated even for isotropic velocity ellipsoid: axial ratios are typically 1:4-1:5 for $\beta=0$ and the NFW profile.

\begin{figure}
\includegraphics[width=0.5\textwidth]{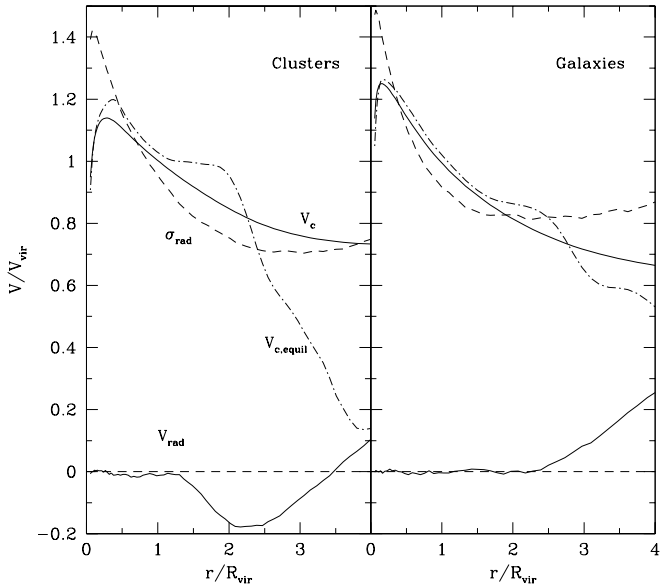}
\caption{Different velocity components for cluster-size haloes (left panel $M_{\rm vir}\approx 2\times 10^{14}h^{-1}{\rm M}_{\sun}$) and galaxy-size haloes (right panel $M_{\rm vir}\approx 10^{12}h^{-1}{\rm M}_{\sun}$). The dot-dashed curves shows the predictions $v_{c, equil}$ of stationary Jeans equation (\ref{eqn:jeans}). The stationary solution closely follows the real circular velocity up to $(2-2.5)R_{\rm vir}$ for galaxy-size haloes. It falls below $v_c$ at larger distances indicating significant non-stationary effects. The situation with clusters is different: the central virialized region is surrounded by a shell where virialization is happening ($v_{c, equil}> v_{c}$) followed by the region where $v_{c, equil}< v_{c}$. See text for details.}
\label{fig:Jeans}
\end{figure}

Why small haloes do not have a turn-around radius followed by an infall region? The top-hat model of non-linear evolution is supposed to provide a reasonable description of what happens at large distances. Yet, it seems to be failing qualitatively for galaxy-size haloes, which do not have infall region and even show an outflow. Haloes are complex systems: they are not spherically symmetric and may have substantial substructure. Still, the spherically symmetric non-stationary Jeans equation is an undeniably significant improvement as compared with the top-hat model. The Jeans equation provides a simple approximation for the haloes. Assuming spherical symmetry the non-stationary Jeans equation can be written in the following form:

\begin{equation}
\frac{r}{\rho}\frac{d(\rho v_r)}{dt} +\sigma_r^2\left[\frac{d\ln(\rho\sigma_r^2)}{d\ln r}+2\beta\right]
       =v^2_c(r),
\label{eq:Jeans}
\end{equation}

\noindent where $v_r$ is the radial velocity and $v^2_c(r)=GM(<r)/r$. When the rms velocities $\sigma_r$ are negligible, the equation (\ref{eq:Jeans}) is the normal equation of motion. Assuming a small initial perturbation and the Hubble flow, we recover the top-hat model. Yet, the haloes do not have small rms velocities even well outside the formal virial radius. Figure~\ref{fig:Jeans} illustrates the point. Even for the massive haloes, which typically dominate their environment, the rms velocities are relatively large. The effect is even stronger for less massive haloes. Another important component is the density gradient. Figure~\ref{fig:dens} shows that the density gradients exist at very large distances: at many virial radii. At $z=2$ the density was on average steeply declining well outside its virial radius.

The left-hand-side of the Jeans equation has two terms. The first term describes non-stationary effects. It is the dominant term at early stages of evolution. The second term is related with the pressure gradient. It dominates in the virial part of the halo. If the system is stationary, the second term must be equal to the square of the circular velocity in the right-hand-side of the Jeans equation. We can call this term the equilibrium circular velocity:
\begin{equation}
v^2_{c. \rm equil}= \sigma_r^2\left[\frac{d\ln(\rho\sigma_r^2)}{d\ln r}+2\beta\right].
\label{eqn:jeans}
\end{equation}
We plot $v_{c, \rm equil}$ in Figure~\ref{fig:Jeans}. For galaxy-size haloes (right panel) in the inner region $r<(2-2.5)R_{\rm vir}$ the term is very close to the right-hand-side of the Jeans equation, which means that the system is nearly in virial equilibrium. At larger distances the term falls short of the circular velocity: the halo is out of equilibrium.

For clusters (the left panel) the equilibrium is only in the central $r<R_{\rm vir}$ region. Just outside of this region at $r=(1-2)R_{\rm vir}$ the equilibrium term $v_{c, \rm equil}$ is substantially above the circular velocity $v_c$. At the same range of radii the infall velocity is clearly detected. The upturn in the $v_{c, \rm equil}$ is the cumulative result of two effects: increase in the slope of the density profile and substantial change in the velocity anisotropy. The same trend exists in the case of small haloes, but it is barely visible. At larger distances $r>2R_{\rm vir}$ the equilibrium term $v_{c, \rm equil}$ falls below the circular velocity.

Our analysis of the Jeans equation shows that the mean radial velocity is a good indicator of the virialization of a halo and that the low-mass haloes are nearly in virial equilibrium well beyond formal $R_{\rm vir}$.
 
\section{Evolution and scaling of halo masses}
\label{sec:scaling}
\begin{figure}
\includegraphics[width=0.5\textwidth]{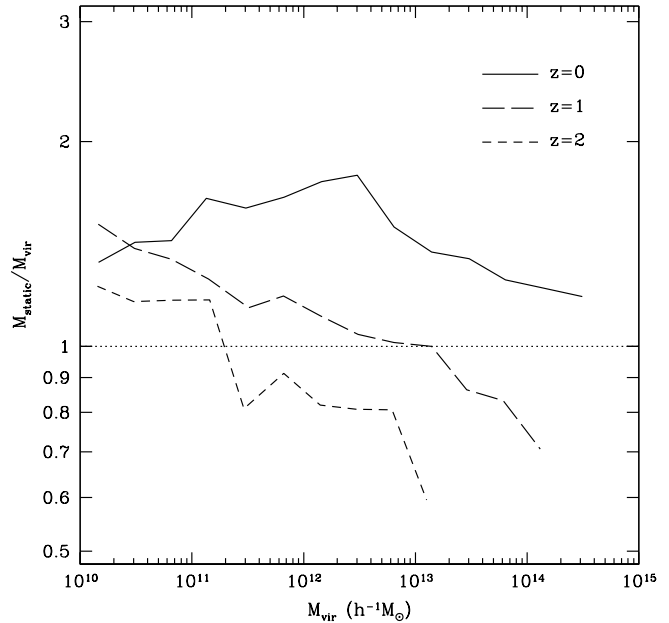}
\caption{The static to virial mass relation from $z=2$ to $z=0$, taking into account every halo at a given redshift. Large haloes having $M_{\rm static} < M_{\rm vir}$ exist at all redshifts, but the transition mass decreases with increasing redshift.}
\label{fig:noscale}
\end{figure}

Dark matter haloes grow by accreting mass from their outskirts. Haloes with small mass grow very slow at late times. Thus, in the past they must have experienced a remarkable infall, which has created extended virialized regions surrounding the haloes at present. In this section we study how this process of building of halo exterior was happening. For this reason we analyze our simulations at different redshifts. We then take the median of different halo properties using the same mass bins as in previous sections with masses ranging from $10^{10}$ to $10^{15}h^{-1}M_{\rm vir}$. As expected, the statistics in each mass bin is smaller at higher redshifts. Using these data we calculate the relation between static and virial mass for $z=0-2$. When doing this analysis we take every halo at a given redshift without considering their relation with the final haloes at $z=0$. The result is shown in Figure~\ref{fig:noscale}, where we plot only the mass bins with enough statistics. At $z=0$ the $M_{\rm static}/M_{\rm vir}$ ratio increases with mass. Then it reaches maximum and starts to decline at large masses. Because of mass resolution we do not see the rising part of $M_{\rm static}/M_{\rm vir}$ ratio at large redshifts, but the declining branch of the curve is better resolved, and we clearly see that the ratio gets below unity at large masses.

It is suggesting that the relation between static and virial mass is essentially scale-free for different redshifts up to $z=2$. In order to test this hypothesis, we approximate the declining part of this relation by a power law $M_{\rm static}/M_{\rm vir}=C M_{\rm vir}^{\alpha}$, and then find the crossing point: $M_{\rm static}=M_{\rm vir}$. This solution $M_0=M_0(z)$ allows us to scale the static to virial mass relation for different redshifts. The results clearly overlap, indicating that there a universal ($z-$independent) static to virial mass relation shown in Figure~\ref{fig:scaled}. This plot shows that the static to virial mass ratio is well fitted by two power-law with different slopes:

\begin{figure}
\includegraphics[width=0.5\textwidth]{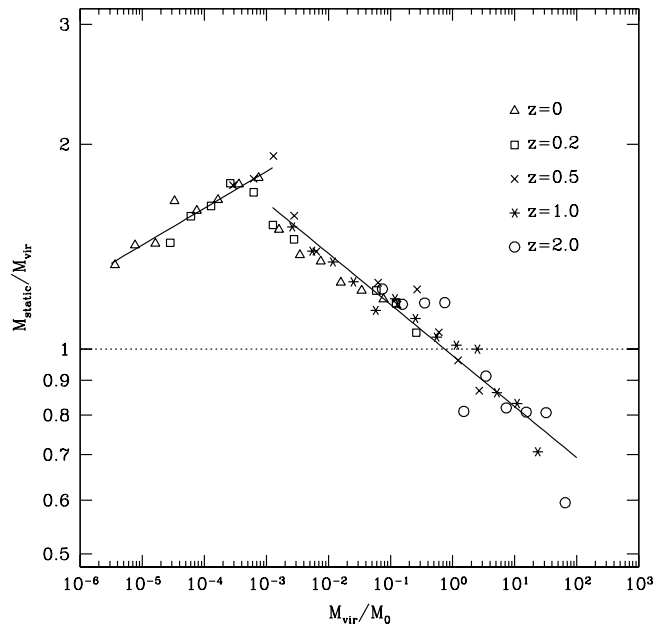}
\caption{Same as Figure~\ref{fig:noscale}, but scaled to the mass $M_0$ in which the fit of the smooth decline in $M_{\rm static}/M_{\rm vir}$ to a power law is equal to one. The time dependence of the $M_{\rm static}$--$M_{\rm vir}$ relation is encoded in the mass scale $M_0$. The solid line represents a fit to a power law at both sides of the transition mass, see equation (\ref{eqn:fit}).}
\label{fig:scaled}
\end{figure}

\begin{equation}
\frac{M_{\rm static}}{M_{\rm vir}}=\left\lbrace \begin{array}{crr}
\mathbf{10^{+0.423 \pm 0.021}}& x^{+0.054 \pm 0.005} & x\lesssim 10^{-3}  \\
\mathbf{10^{-0.010 \pm 0.005}}& x^{-0.075 \pm 0.004} & x\gtrsim  10^{-3}  \\
\end{array}
\right.
\label{eqn:fit}
\end{equation}
\noindent where $x\equiv M_{\rm vir}/M_0$. This relation allows us to convert the virial mass, which is easier to be computed accurately in numerical simulations, to the static mass. In order to have a complete description of this relation for any redshift, it remains to be determined the redshift dependence of the mass scale $M_0(z)$. Although this dependence may suffer from the scatter in the determination of the static mass, it seems clear from Figure~\ref{fig:m0evol} that there is an approximate power-law dependence $M_0(a)=M_0(a=1)a^{\gamma}$, from $z=2$ to $z=0$. The best fit to our data\footnote{We want to emphasize that halo virial mass is defined here with a overdensity criterion which evolves with time, as detailed in Section~\ref{sec:simul}. The choice of non-evolving overdensity criterion will imply a change in the $M_0(z)$ relation.} is given by:

\begin{equation}
M_0 \simeq (2.8\pm 0.8)\times10^{15}a^{8.9\pm0.4}h^{-1}{\rm M}_{\sun}
\label{eqn:m0evol}
\end{equation}

\begin{figure}
\includegraphics[width=0.5\textwidth]{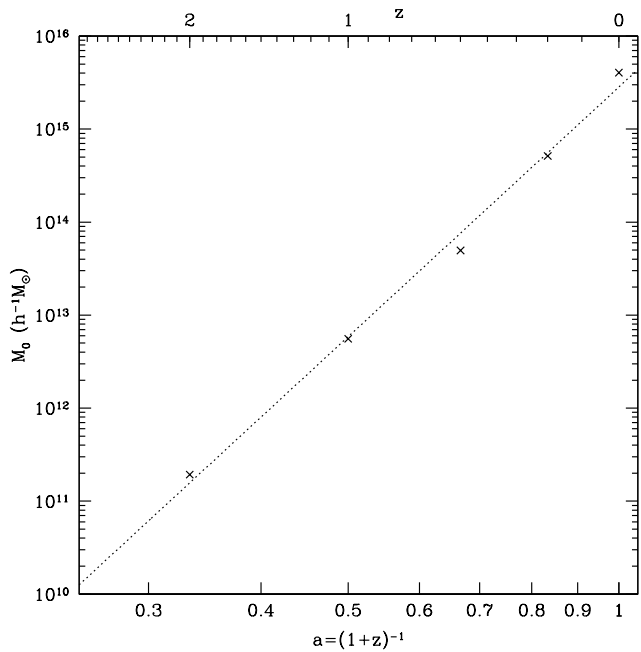}
\caption{The evolution of the mass scale $M_0$. This behaviour is approximately fitted by a power-law,$ M_0\simeq 2.8\times10^{15} a^{8.9} h^{-1}{\rm M}_{\sun}$ (see text for details).}
\label{fig:m0evol}
\end{figure}

It is worth of pointing that the absence of explicit redshift dependence in the relation between static and virial mass when the variable $M_{\rm vir}/M_0$ is used, allows an analytic and straightforward derivation of the relation between both mass accretions, i.e. taking the logarithmic derivative in equation~(\ref{eqn:fit}) and using equation~(\ref{eqn:m0evol}) we obtain: 
\begin{equation}
\frac{\dot{M}_{\rm static}}{M_{\rm static}}=\left(1+\alpha\right)\frac{\dot{M}_{\rm vir}}{M_{\rm vir}}-\frac{\alpha\gamma}{a},
\label{eqn:model}
\end{equation}
where the dot refers to the derivative with respect to the scale factor $a$. However, this equation is no longer valid near the transition mass, where this derivative is not finite.

Despite the mass scale $M_0$ is not motivated by the physics of the formation of dark matter haloes (as opposed to e.g. $M_{*}$), it is the best way we found to encode the redshift dependence of $M_{\rm static}/M_{\rm vir}$. We have also explored the relation of this ratio with $\nu=\delta_c/\sigma(M,z)$. However, we found a wider scatter for a given value of $\nu$, with lower values of the ratio for increasing redshift, in particular for $\nu>1$.

\section{Mass function}
\label{sec:mf}
\begin{figure}
\includegraphics[width=0.5\textwidth]{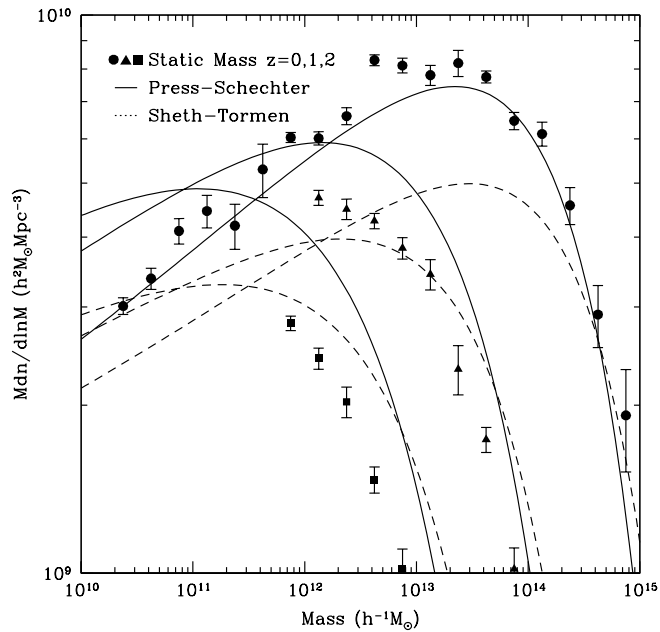}
\caption{The static mass function obtained from individual haloes using equation (\ref{eqn:fit}) (solid circles for $z=0$, triangles for $z=1$ and squares for $z=2$). Error bars show the Poissonian error only (not including the uncertainty in the $M_{\rm static}-M_{\rm vir}$ relation). Solid and dashed lines show the PS and ST mass function, respectively. Surprisingly, the PS mass function provides a very good fit to our data at $z=0$. On the other hand, the mass function at higher redshifts deviates from the PS analytic model.}
\label{fig:smf}
\end{figure}

The relation between the static and the virial mass is not monotonic, as we have shown before. This implies a change in the shape and in the amplitude of the distribution of the number density of haloes with a given mass, i.e. the mass function. As the virial mass is calculated accurately for all the haloes in our simulations, we may also obtain their static mass directly from the conversion formula given in equation~(\ref{eqn:fit}). This allows us to estimate the static mass function of dark matter haloes and compare it to the virial mass function. We obtain very similar results taking the virial mass bins in Figure~\ref{fig:scaled} and then shifting them according to their corresponding value of the ratio $M_{static}/M_{vir}$. The results are presented in Figure~\ref{fig:smf}. In spite of the fact that the mass function of \citet{She99} provides a very good fit for the virial mass (as shown in Figure~\ref{fig:mf} for the case of $z=0$), we find that this analytic model is not a good description for the static mass function. Moreover, some authors using different definitions of virialized halo have already shown the presence of significant departures from the Sheth \& Tormen function, see e.g. \citet{Mac03}.

Surprisingly, the mass function of \citet{Pre74} approximates really well our data at $z=0$ in the mass range under study. In order to test robustness of our estimates of the static mass function, we changed different parameters (e.g., changed the threshold to $0.1V_{\rm vir}$ in algorithm of detecting the static radius). Results are stable. We leave for future the investigation of the static mass function at its high mass end \citep{Bet06}, because the poor statistics for the highest mass bins here prevents us from drawing any conclusion about it. The redshift evolution of the mass function is also shown in Figure~\ref{fig:smf}. We realize that the static mass function deviates from the corresponding Press \& Schechter function at $z=1$ and $z=2$ so it is no longer described by this model. None the less, it is interesting that we can still recover the agreement with Press \& Schechter by an overall increase in the static mass. This shift could be accounted for by the scatter in the $M_0(z)$ relation, which is very similar to this deviation. However it is not clear whether this deviation is an artifact due to resolution effects, so this remains as an open issue.

\section{Evolution of the Major halo Progenitor}
\label{sec:evolution}
We are now in a position to study the evolution of the mass, i.e. the mass accretion history, of dark matter haloes in cosmological simulations. So, we need to track the set of haloes identified by BDM at $z=0$ back in time and then measure their static mass. Here we use a method similar to that in \citet{Asc07}. First of all, we pick up a sample of few hundreds of particles around the density maximum of every halo at $z=0$. These particles are then tracked back to the highest redshift available in our cosmological simulation. Although it may occur that these particles no longer belong to the same object at that redshift, we take the particle which has a minimum sum of squared distances to the others in the sample, as an initial guess for the centre of the protohalo. Finally, we calculate the centre of mass around this position iteratively using the whole set of particles in the simulation box. After few iterations, we obtain the centre of the primordial peak of density which corresponds to our halo at $z=0$. We then label some particles around this centre. This new set of particles will help us, by calculation of their centre of mass, to identify at every redshift the position of the halo which encloses them. Hence this halo, as the offspring of the main primordial peak corresponding to the halo at present, will be referred to as the halo major progenitor.

In our analysis we study the evolution of the progenitors from three different mass bins at $z=0$: low-mass haloes ($10^{10.0}$--$10^{10.5} h^{-1}{\rm M}_{\sun}$), galactic-size haloes ($10^{12.0}$--$10^{12.5} h^{-1}{\rm M}_{\sun}$) and cluster-size haloes ($10^{13.5}$--$10^{14.0} h^{-1}{\rm M}_{\sun}$). These ranges are selected in order to study the dependence of mass on our results with very good statistics: each bin contains more than a thousand of haloes. This allows us to calculate with good accuracy the median profiles for different physical quantities at every redshift from $z=0$ to $z\simeq 2$. We keep this as the maximum redshift analysed here because the median radial profile is no longer well resolved (i.e. with more than 200 particles for the majority of the haloes) in the inner regions for higher redshifts. Thus, we cannot determine the size of the virialized region and hence we cannot go further in redshift.

\begin{figure*}
\begin{center}
\includegraphics[width=0.3\textwidth]{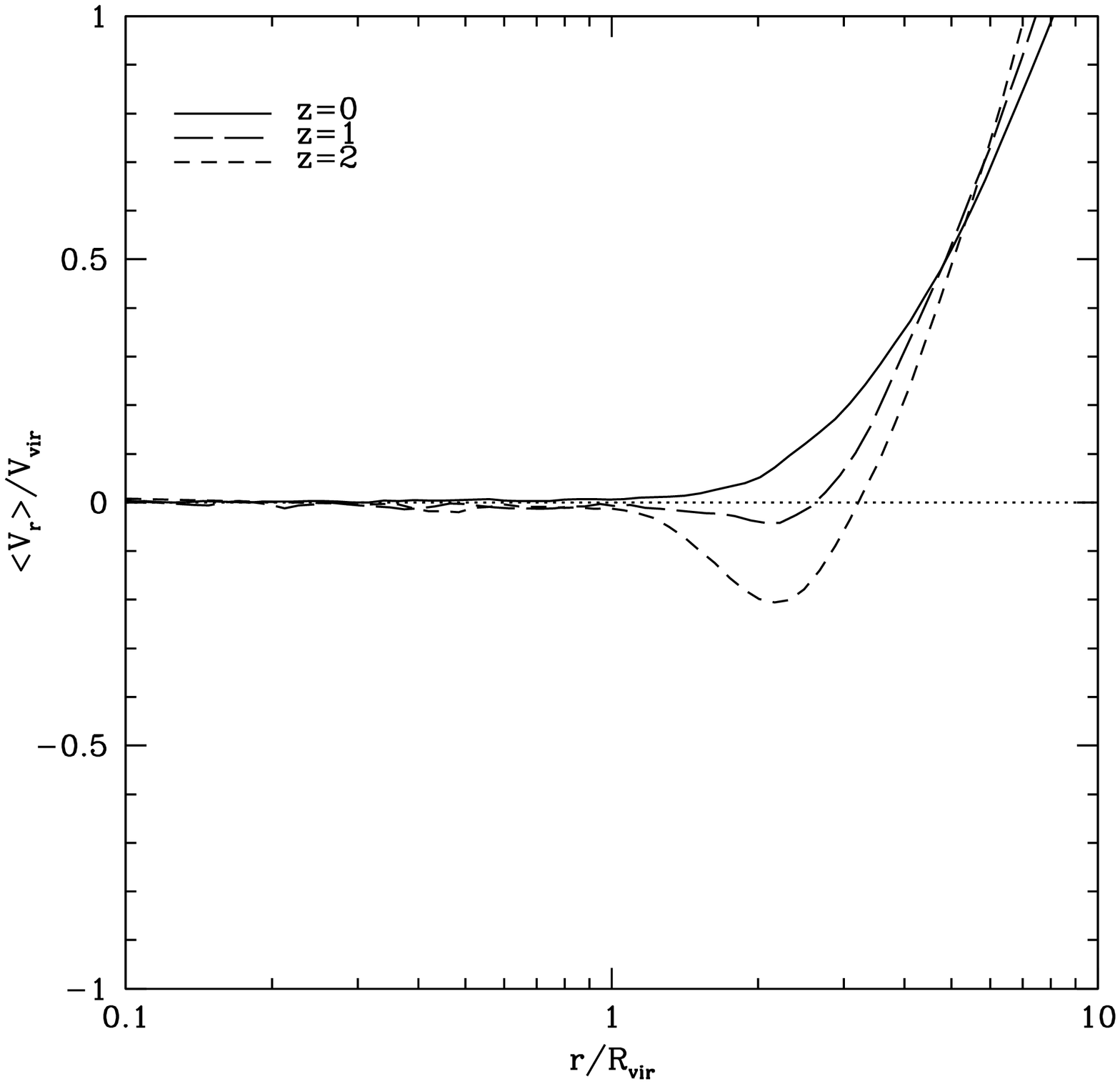}
\includegraphics[width=0.3\textwidth]{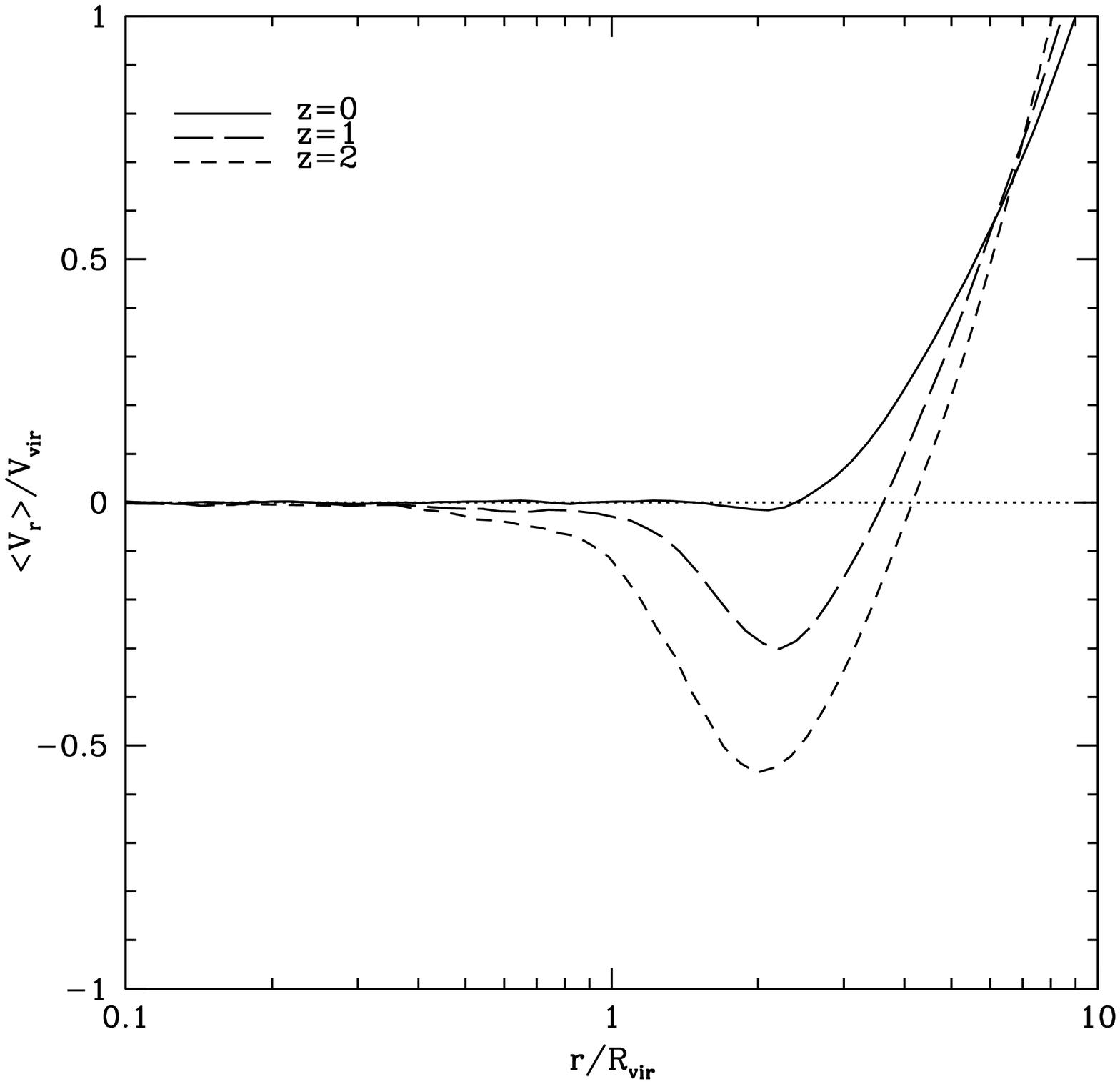}
\includegraphics[width=0.3\textwidth]{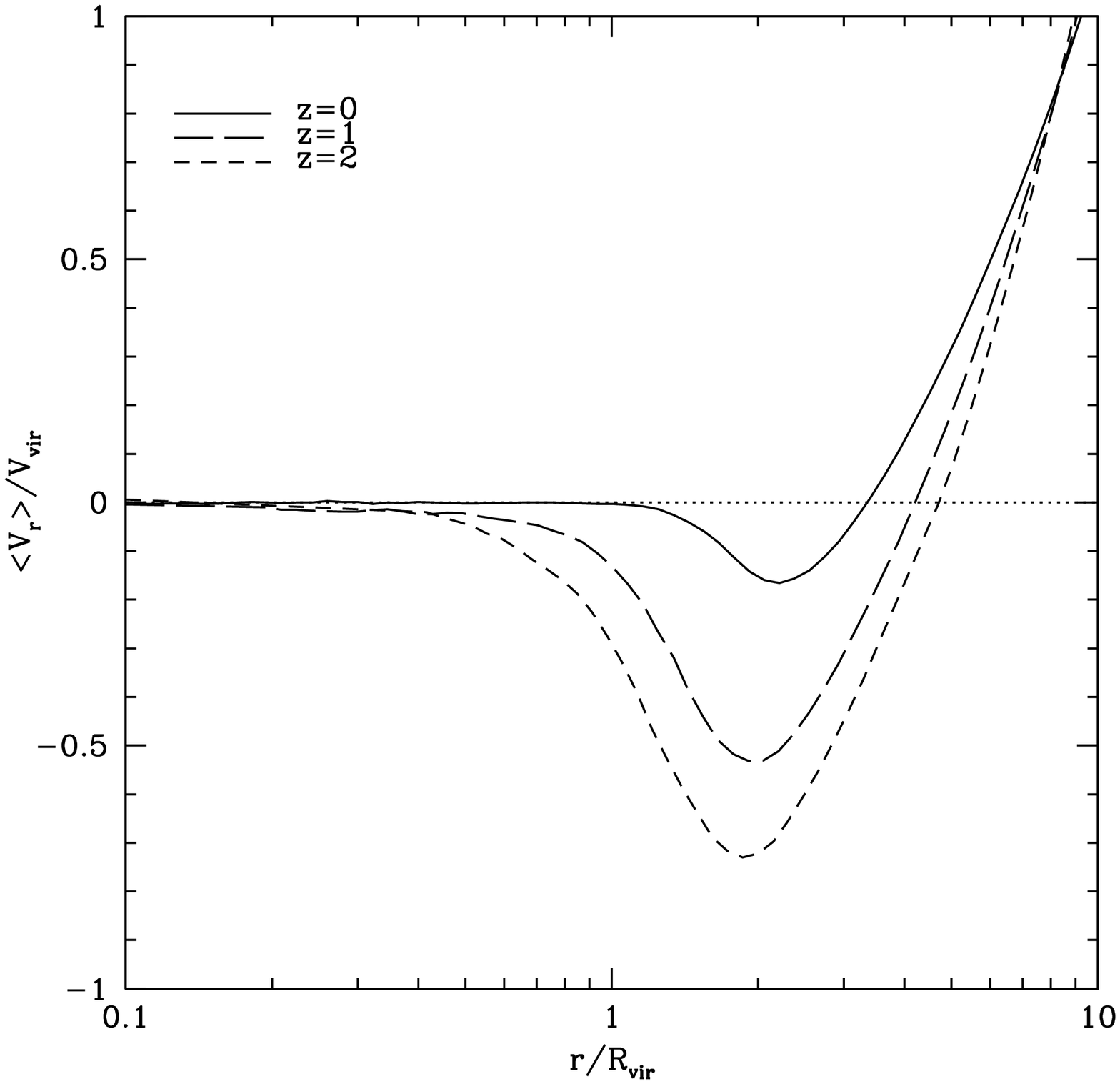}
\end{center}
\caption{The evolution of the mean radial velocity profile from $z=2$ to $z=0$. Left panel: low-mass haloes. Middle panel: galactic-size haloes. Right panel: cluster-size haloes. The sequence followed by low-mass haloes has hardly started by cluster-size haloes.}
\label{fig:vrevol}
\end{figure*}

The evolution of the mean radial velocity profile for the median halo is displayed in Figure~\ref{fig:vrevol}. Here we show that the time sequence suggested in \citet{Bus05} for cluster-size haloes takes place even for low-mass haloes: the mean radial velocity profile experiences infall in their outer parts while forming. Just after the infall is finished, there is an epoch in which some outflow appears, followed by a final profile with an inner virialized region and the Hubble flow\footnote{This final state agrees with the picture shown by other different analyses of halo virialization, in which virial parameters are ill-defined in merging events but there is a general trend for the virial ratio $2E_{\rm kin}/\vert E_{\rm pot}\vert$ to approach unity at present, e.g. \citet{Het06}.}. Moreover, the time scale of this sequence seems to be mass-dependent: the analysis of \citet{Bus05} finds this sequence in the future for cluster-size haloes, whereas low-mass haloes are already in the outflow phase at present. Galaxy-size haloes are now experiencing neither infall nor outflow, so the virialized region is reaching a maximum in units of the virial radius, as already shown in Figure~\ref{fig:ms}.

\begin{figure}
\includegraphics[width=0.5\textwidth]{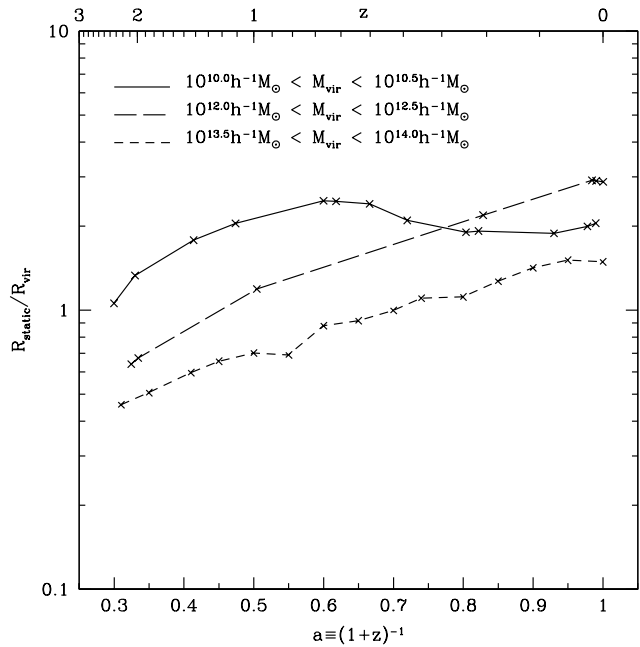}
\caption{The static to virial radius as a function of the redshift. It appears that this ratio behaves differently for each mass bin. The mass bins are the same as those on Figure~\ref{fig:vrevol}. Low-mass haloes have reduced their static region in units of $R_{\rm vir}$ due to recent outflow. On the other hand, galaxy-size haloes are reaching a maximum on this ratio. Cluster-size haloes continue increasing this ratio as the infall is less prominent.}
\label{fig:revol}
\end{figure}

This evolution is reflected in the redshift dependence of the static radius. In Figure~\ref{fig:revol} we illustrate the evolution of the ratio of static to virial radius. It is clear from this plot that low-mass haloes have already formed (the infall stopped when the maximum ratio took place) and they are even losing some of their virialized mass at present. Galaxy-size haloes are reaching now this maximum, but for cluster-size haloes the infall has not even finished.

\begin{figure}
\includegraphics[width=0.5\textwidth]{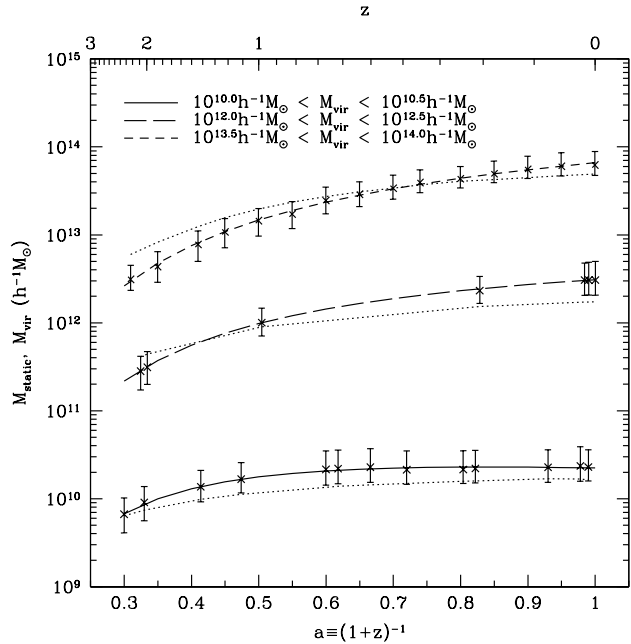}
\caption{The mass accretion history of dark matter haloes. The evolution of the static mass is indicated by the crosses and the error bars, with their corresponding fit to equation (\ref{eqn:modfit}) (solid, long-dashed and dashed lines for low-mass, galaxy-size and cluster-size haloes, respectively), while the dotted line shows the evolution of the virial mass, according to the parametrization by \citet{Wec02}. The mass bins are the same as those on Figure~\ref{fig:vrevol}. The static mass inside low-mass haloes is nearly constant from $z=0.5$ to $z=0$. On the other hand, galaxy-size haloes increase their static mass even steeper than their virial mass, due to relaxation around haloes. Clusters are accreting mass, but also incorporating some of the recently virialized mass from their surroundings.}
\label{fig:mah}
\end{figure}

The static mass accretion history is presented in Figure~\ref{fig:mah}. The solid, long-dashed and short-dashed lines show the evolution of the static mass of the median halo for three different mass bins, while the dotted line corresponds to their virial mass. The error bars are estimated using the halo-to-halo variation, calculating the mass inside the median virial radius for each dark matter halo, and taking the central 68 per cent of these values. Neither the uncertainty in the virial radius nor the scatter in the static vs. virial mass relation were taken into account. Despite of the scatter, some conclusions may be drawn from this plot. The recent assembly history is very mass-dependent: low mass haloes hardly experienced a change in their mass since $z=1$, and their virial mass is under-predicting the average amount of matter inside them since $z\simeq 2$. On the other hand, galactic-size haloes have been increasing their static mass up to the present epoch, most likely due to halo relaxation. This process is responsible of the largest difference between the static and the virial mass at $z=0$: the mass inside the virialized region is about a factor of two larger than the virial mass. Finally, cluster-size haloes are still increasing their mass at present, but the difference with respect to the virial mass is not so large. In fact, around $z=0.5$ both static and virial mass should have been very similar. On the contrary, it is likely that they differ significantly at higher redshift.

We find for our median profiles that the function proposed in equation (3) in \citet{Wec02} for the virial mass, $M_{\rm vir}(a)=M_{\rm vir}(a=1)\exp\left(-\delta z\right)$ provides an excellent fit for the accretion history of the virial mass from $z=2$ to $z=0$. However, if we include this relation in our model given by equation (\ref{eqn:model}), and solving the differential equation we obtain:
\begin{equation}
\frac{M_{\rm static}(a)}{M_{\rm static}(a=1)}=\frac{\exp^{-\left(1+\alpha\right)\delta z}}{a^{\alpha\gamma}}
\end{equation}
which fails to reproduce the evolution of the static mass when the discontinuity of the static vs virial mass relation takes place. This is reasonable as equation (\ref{eqn:model}) is obtained when all the haloes (not just the progenitors of the haloes at $z=0$) are taken into account. None the less, this starting point suggests the functional dependence of this evolution, so that a fit of the data to:
\begin{equation}
M_{\rm static}(z)=M_{\rm static}(z=0)(1+z)^{-\beta}e^{-\alpha z}
\label{eqn:modfit}
\end{equation}
where both $\alpha$ and $\beta$ are free parameters, provides a better description of the data. This function is very similar to that corresponding to the virial mass, but there is an extra factor which enhances the fit, in particular for low-mass haloes.

\begin{figure}
\includegraphics[width=0.5\textwidth]{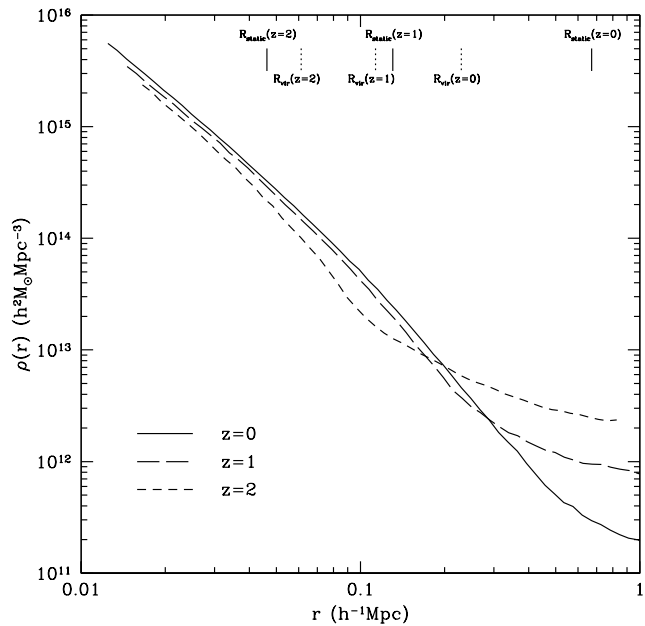}
\caption{The evolution of the median density profile from $z=2$ to $z=0$ for galactic-size haloes. As shown in Fig.~\ref{fig:dens} for individual haloes, there has been very little change in this profile from $z=1$ up to the present epoch. The vertical dotted lines marks the virial radii at $z=2$ (left), $z=1$ (middle) and $z=0$ (right). Static radii are shown with vertical solid lines. The growth of the static radius in Milky-Way size haloes from $z=1$ to $z=0$ is not due to accretion but virialization of the surrounding regions of the halo.}
\label{fig:rho}
\end{figure}

As we pointed out above, the recent increase in the mass of galactic-size haloes is mainly due to virialization of mass surrounding them. In Figure~\ref{fig:rho} we display the evolution of the density profile for the median galactic-size halo, which is similar to the profile for individual haloes as shown in Figure~\ref{fig:dens}. Only small variations in the density profile occur for the median galactic halo from $z=1$ to $z=0$, which correspond to a small variation of the mass inside a physical radius. However, these haloes experience an important growth of their static mass due to the expansion of the virialized region, as it is clear from Figure~\ref{fig:vrevol}.

\section{Discussion}
\label{sec:discussion}
A more realistic (physical) estimate of the halo mass might be crucial for some areas in cosmology which are sensitive to accurate values of the mass of dark matter haloes. For example, our results may be relevant for understanding the galaxy formation processes that happen inside dark matter haloes. Indeed, semi-analytical models of galaxy formation (e.g. \citealt{Cro06}) might need to assess the impact of their predictions when the virial mass is no longer used. If the static mass and its evolution are used, one may get a more realistic picture of the structure formation process drawn from cosmological simulations. Moreover, recipes for star formation, AGN and supernova feedback and merging of galaxies are often taken from quantities of their corresponding dark matter haloes, such as their mass. Thus, the conclusions (e.g. colour-magnitude diagram, evolution of the stellar mass, etc.) from the analysis of these simulations using semi-analytical models of galaxy formation may be biased due to an underestimation of the mass associated to a halo if the former virial mass is considered, especially for the galaxy-size haloes such as the host of the Milky Way as we have reported in this work.

The precise determination of the halo mass function, in terms of the static mass, can also be very relevant. Mass function of cluster-size haloes must be predicted with special accuracy because the number density of galaxy clusters is a probe of fundamental cosmological parameters, such as dark energy. This is very important since the evolution of the mass function of galaxy clusters is taken as an almost direct measurement of the equation of state of dark energy $\omega=P/ \rho$ \citep{Abb05} and any statistically significant deviations in the measurements of the mass function of clusters from the predictions of the $\Lambda$CDM model, may be and likely will be interpreted as deviations of the dark energy from a simple cosmological constant.

Other possible observational implications of this work also deserve further investigation. It is noteworthy that the transition mass of $\simeq 5\times10^{12}h^{-1}{\rm M}_{\sun}$, that separates the infall and outflow regimes, is similar to the mass scale found by \citet{Dek06}. \citet{Dek06} argue that below $M\lesssim 10^{12}{\rm M}_{\sun}$ the efficient early star formation may occur due to cold flows, as thermal pressure is not large enough to avoid the gas accretion from the outer gas reservoir to the inner core of the halo. On larger scales the heating due to virial shocks may shutdown the star formation. This different behaviour of the gas around dark matter haloes in simulations above and below this mass scale was proposed by \citet{Dek06} as an explanation for the bimodality observed in many properties of galaxies such as the bimodality seen in the colour distribution of SDSS field galaxies. Thus, a given velocity pattern of infall or outflow in a dark matter halo of mass above and below $10^{12}h^{-1}{\rm M}_{\sun}$ may be correlated with the fact that the galaxy within the halo belongs to the blue cloud or the red colour sequence, respectively. Yet, this is only an speculation which remains to be investigated. However, the mass scale for transition between infall and outflow is the same as the mass scale at which baryonic matter cannot penetrate the inner parts of the halo. At first glance this is just a matter of coincidence, as for higher redshifts the transition mass would be well below the mass scale of \citet{Dek06}. Still, it is an interesting question whether the haloes experiencing infall of the dark matter also host galaxies, which grow fast by accreting baryons in their central regions.

Such correlations between galaxies and their host haloes could unveil some of the relations between dark and baryonic matter, and hence cast some light in the evolution of galaxies in the expanding Universe.

\section{Conclusions}
\label{sec:conclusion}

In this paper we use a set of high resolution $N$--body cosmological simulations for the analysis of the mass inside the region of dark matter haloes with no net infall or outflow velocities, i.e. the static region. Haloes show a typical radial velocity pattern which depends on their mass: low-mass haloes tend to display a region with outflow in their surroundings, while cluster-size haloes show a prominent infall velocity pattern. Galaxy-size haloes show an intermediate pattern with a sharper transition between the static region and the Hubble flow. We find that the former virial radius tends to underestimate the size of the region with zero mean radial velocity for haloes with masses $10^{10}h^{-1}{\rm M}_{\sun}<M_{\rm vir}\lesssim 10^{14}h^{-1}{\rm M}_{\sun}$ at $z=0$. The mass inside this static region can be about a factor of two larger than the former virial mass, when the threshold defining the static radius is $5$ per cent of the virial velocity $V_{\rm vir} =\sqrt(GM_{\rm vir}/R_{\rm vir})$. Lower values of the threshold ($\lesssim 1$ per cent) may lower this factor. However, we observe a clear trend in the $M_{\rm static}/M_{\rm vir}$ vs $M_{\rm vir}$ relation which remains unchanged with this threshold: for low-mass haloes, this ratio is an increasing function of $M_{\rm vir}$ due to reduction in the outflow. On the other hand, high-mass haloes show a larger infall for larger $M_{\rm vir}$, making this ratio to decrease. The maximum occurs at $M_{\rm vir}\simeq 5\times 10^{12}h^{-1}{\rm M}_{\sun}$, where the size of the static region is the largest in units of $R_{\rm vir}$.

The mass function of objects whose mass is defined in the way presented here resembles that of Press \& Schechter in the range of mass we have studied, but only at $z=0$. At higher redshifts the static mass function deviates significantly from it. This disagreement could be accounted for by the scatter in $M_0(z)$, although it seems more likely that the resemblance with Press \& Schechter at $z=0$ is due to coincidence (in both shape and amplitude to within $\sim 20\%$) of the ratio $f_{PS}/f_{ST}$ with $M_{\rm static}/M_{\rm vir}$. Whatever the case may be, this resemblance is theoretically unexpected because the Sheth \& Tormen function, which fits very well the mass function for virial masses, is derived from a more realistic approach (ellipsoidal collapse). In any case it seems clear that when the virial mass is used, the number density of dark matter haloes of a given mass at $z=0$ might be underestimated in the mass range $10^{10}h^{-1}{\rm M}_{\sun}<M_{\rm vir}\lesssim 10^{14}h^{-1}{\rm M}_{\sun}$.

The redshift evolution of the $M_{\rm static}$--$M_{\rm vir}$ relation turns out to be very weakly redshift-dependent from $z=2$ to $z=0$ when appropriate variables are used. This dependence on the redshift is encoded as an evolving mass scale $M_0$ which indicates the approximate mass at which $M_{\rm static}=M_{\rm vir}$ in the declining part of the relation. By doing this, it is straightforward to derive from this relation a differential equation for the evolution of the static mass, provided that the evolution of $M_{\rm vir}$ and of $M_0$ are already known. As we have shown in this paper, the evolution of the mass scale has an acceptable fit to a power law: $ M_0\simeq 2.8\times10^{15} a^{8.9} h^{-1}{\rm M}_{\sun}$ . On the other hand, the evolution of the virial mass $M_{\rm vir}$ of the major progenitor is well described by an exponential law in $z$, as claimed by \citet{Wec02}. Once we have fixed the evolution of $M_{\rm vir}$ and of $M_0$, we solve this differential equation and we find a simple model for the evolution of the static mass of the major progenitor, i.e. $M_{\rm static}(z) =M_{\rm static}(z=0)(1+z)^{-\beta}e^{-\alpha z}$, which turns out to be a very good fit to our data. In any case, there is a different behaviour for the evolution of the static mass in different mass bins. The static mass of low mass haloes is nearly constant from $z=0.5$ to $z=0$. Galaxy-size haloes keep growing at present, but the reason here is not the decreasing background density of the expanding Universe. Instead, the relaxation of the matter in the surroundings of these haloes is incorporating mass at even a higher rate than the unphysical growth of the virial mass. The same stands for clusters, although the static and virial mass are not very different at present.

The mean radial velocity profile (averaged over a halo mass bin) evolves in the way found by \citet{Bus05} for cluster-size haloes: while the halo is forming, it accretes mass from its surroundings via infall, which is being reduced with time. This is followed by a period of outflow of unbound matter, and a final configuration thereafter consisting in a static region whose size is independent of time. Although the simulation used in that paper, was evolved until $a=100$, and our halo sample was instead tracked up to redshift $z=0$, we find a situation that is consistent with this picture. Moreover, we find that the timescale of this sequence is different for each mass bin: low-mass haloes at present are in the outflow phase, but cluster-size haloes are still in their infall phase. Interestingly, galactic-size haloes are experiencing a transition (infall to outflow) epoch at $z=0$.

A.J.C., F.P. thank the Spanish MEC under grant PNAYA 2005-07789 for their support. A. K. acknowledges support from NASA and NSF grants to NMSU. A.J.C. acknowledges the financial support of the MEC through Spanish grant FPU AP2005-1826. We thank Juan Betancort-Rijo for valuable discussion and useful comments in this work, and Douglas Rudd for suggesting some points which improved the text. We do appreciate the useful feedback from the anonymous referee. A.J.C. also thanks the Helmholtz Institute for Supercomputational Physics for their outstanding Summerschool 2006 in Supercomputational Cosmology. Computer simulations were done at the LRZ Munich, NIC Julich, and NASA Ames.

\bibliography{mycites}

\begin{thebibliography}{}

\bibitem[\protect\citeauthoryear{{Ascasibar}, {Hoffman} \&
  {Gottl{\"o}ber}}{{Ascasibar} et~al.}{2007}]{Asc07}
{Ascasibar} Y.,  {Hoffman} Y.,    {Gottl{\"o}ber} S.,  2007, \mnras, 376, 393

\bibitem[\protect\citeauthoryear{{Betancort-Rijo} \&
  {Montero-Dorta}}{{Betancort-Rijo} \& {Montero-Dorta}}{2006}]{Bet06}
{Betancort-Rijo} J.~E.,  {Montero-Dorta} A.~D.,  2006, \apjl, 650, L95

\bibitem[\protect\citeauthoryear{{Blumenthal}, {Faber}, {Primack} \&
  {Rees}}{{Blumenthal} et~al.}{1984}]{Blu84}
{Blumenthal} G.~R.,  {Faber} S.~M.,  {Primack} J.~R.,    {Rees} M.~J.,  1984,
  \nat, 311, 517

\bibitem[\protect\citeauthoryear{{Bryan} \& {Norman}}{{Bryan} \&
  {Norman}}{1998}]{Bry98}
{Bryan} G.~L.,  {Norman} M.~L.,  1998, \apj, 495, 80

\bibitem[\protect\citeauthoryear{{Bullock}, {Dekel}, {Kolatt}, {Kravtsov},
  {Klypin}, {Porciani} \& {Primack}}{{Bullock} et~al.}{2001}]{Bul01b}
{Bullock} J.~S.,  {Dekel} A.,  {Kolatt} T.~S.,  {Kravtsov} A.~V.,  {Klypin}
  A.~A.,  {Porciani} C.,    {Primack} J.~R.,  2001, \apj, 555, 240

\bibitem[\protect\citeauthoryear{{Bullock}, {Kolatt}, {Sigad}, {Somerville},
  {Kravtsov}, {Klypin}, {Primack} \& {Dekel}}{{Bullock} et~al.}{2001}]{Bul01}
{Bullock} J.~S.,  {Kolatt} T.~S.,  {Sigad} Y.,  {Somerville} R.~S.,  {Kravtsov}
  A.~V.,  {Klypin} A.~A.,  {Primack} J.~R.,    {Dekel} A.,  2001, \mnras, 321,
  559

\bibitem[\protect\citeauthoryear{{Busha}, {Adams}, {Wechsler} \&
  {Evrard}}{{Busha} et~al.}{2003}]{Bus03}
{Busha} M.~T.,  {Adams} F.~C.,  {Wechsler} R.~H.,    {Evrard} A.~E.,  2003,
  \apj, 596, 713

\bibitem[\protect\citeauthoryear{{Busha}, {Evrard}, {Adams} \&
  {Wechsler}}{{Busha} et~al.}{2005}]{Bus05}
{Busha} M.~T.,  {Evrard} A.~E.,  {Adams} F.~C.,    {Wechsler} R.~H.,  2005,
  \mnras, 363, L11

\bibitem[\protect\citeauthoryear{{Cole} \& {Lacey}}{{Cole} \&
  {Lacey}}{1996}]{Col96}
{Cole} S.,  {Lacey} C.,  1996, \mnras, 281, 716

\bibitem[\protect\citeauthoryear{{Croton}, {Springel}, {White}, {De Lucia},
  {Frenk}, {Gao}, {Jenkins}, {Kauffmann}, {Navarro} \& {Yoshida}}{{Croton}
  et~al.}{2006}]{Cro06}
{Croton} D.~J.,  {Springel} V.,  {White} S.~D.~M.,  {De Lucia} G.,  {Frenk}
  C.~S.,  {Gao} L.,  {Jenkins} A.,  {Kauffmann} G.,  {Navarro} J.~F.,
  {Yoshida} N.,  2006, \mnras, 365, 11

\bibitem[\protect\citeauthoryear{{Dekel} \& {Birnboim}}{{Dekel} \&
  {Birnboim}}{2006}]{Dek06}
{Dekel} A.,  {Birnboim} Y.,  2006, \mnras, 368, 2

\bibitem[\protect\citeauthoryear{{Diemand}, {Kuhlen} \& {Madau}}{{Diemand}
  et~al.}{2007}]{Die07}
{Diemand} J.,  {Kuhlen} M.,    {Madau} P.,  2007, \apj, 667, 859

\bibitem[\protect\citeauthoryear{{D'Onghia} \& {Navarro}}{{D'Onghia} \&
  {Navarro}}{2007}]{Don07}
{D'Onghia} E.,  {Navarro} J.~F.,  2007, \mnras, 380, L58

\bibitem[\protect\citeauthoryear{{Gladders}, {Yee}, {Majumdar}, {Barrientos},
  {Hoekstra}, {Hall} \& {Infante}}{{Gladders} et~al.}{2007}]{Gla07}
{Gladders} M.~D.,  {Yee} H.~K.~C.,  {Majumdar} S.,  {Barrientos} L.~F.,
  {Hoekstra} H.,  {Hall} P.~B.,    {Infante} L.,  2007, \apj, 655, 128

\bibitem[\protect\citeauthoryear{{Gunn}}{{Gunn}}{1977}]{Gun77}
{Gunn} J.~E.,  1977, \apj, 218, 592

\bibitem[\protect\citeauthoryear{{Gunn} \& {Gott}}{{Gunn} \&
  {Gott}}{1972}]{Gun72}
{Gunn} J.~E.,  {Gott} J.~R.~I.,  1972, \apj, 176, 1

\bibitem[\protect\citeauthoryear{{Guzik} \& {Seljak}}{{Guzik} \&
  {Seljak}}{2002}]{Guz02}
{Guzik} J.,  {Seljak} U.,  2002, \mnras, 335, 311

\bibitem[\protect\citeauthoryear{{Hetznecker} \& {Burkert}}{{Hetznecker} \&
  {Burkert}}{2006}]{Het06}
{Hetznecker} H.,  {Burkert} A.,  2006, \mnras, 370, 1905

\bibitem[\protect\citeauthoryear{{Hoekstra}, {Yee} \& {Gladders}}{{Hoekstra}
  et~al.}{2004}]{Hoe04}
{Hoekstra} H.,  {Yee} H.~K.~C.,    {Gladders} M.~D.,  2004, \apj, 606, 67

\bibitem[\protect\citeauthoryear{{Klypin}, {Gottl{\"o}ber}, {Kravtsov} \&
  {Khokhlov}}{{Klypin} et~al.}{1999}]{Kly99}
{Klypin} A.,  {Gottl{\"o}ber} S.,  {Kravtsov} A.~V.,    {Khokhlov} A.~M.,
  1999, \apj, 516, 530

\bibitem[\protect\citeauthoryear{{Klypin}, {Kravtsov}, {Bullock} \&
  {Primack}}{{Klypin} et~al.}{2001}]{Kly01}
{Klypin} A.,  {Kravtsov} A.~V.,  {Bullock} J.~S.,    {Primack} J.~R.,  2001,
  \apj, 554, 903

\bibitem[\protect\citeauthoryear{{Kneib}, {Hudelot}, {Ellis}, {Treu}, {Smith},
  {Marshall}, {Czoske}, {Smail} \& {Natarajan}}{{Kneib} et~al.}{2003}]{Kne03}
{Kneib} J.-P.,  {Hudelot} P.,  {Ellis} R.~S.,  {Treu} T.,  {Smith} G.~P.,
  {Marshall} P.,  {Czoske} O.,  {Smail} I.,    {Natarajan} P.,  2003, \apj,
  598, 804

\bibitem[\protect\citeauthoryear{{Kravtsov}, {Klypin} \& {Khokhlov}}{{Kravtsov}
  et~al.}{1997}]{Kra97}
{Kravtsov} A.~V.,  {Klypin} A.~A.,    {Khokhlov} A.~M.,  1997, \apjs, 111, 73

\bibitem[\protect\citeauthoryear{{{\L}okas} \& {Mamon}}{{{\L}okas} \&
  {Mamon}}{2001}]{Lok01}
{{\L}okas} E.~L.,  {Mamon} G.~A.,  2001, \mnras, 321, 155

\bibitem[\protect\citeauthoryear{{Macci{\`o}}, {Dutton}, {van den Bosch},
  {Moore}, {Potter} \& {Stadel}}{{Macci{\`o}} et~al.}{2007}]{Mac07}
{Macci{\`o}} A.~V.,  {Dutton} A.~A.,  {van den Bosch} F.~C.,  {Moore} B.,
  {Potter} D.,    {Stadel} J.,  2007, \mnras, 378, 55

\bibitem[\protect\citeauthoryear{{Macci{\`o}}, {Murante} \&
  {Bonometto}}{{Macci{\`o}} et~al.}{2003}]{Mac03}
{Macci{\`o}} A.~V.,  {Murante} G.,    {Bonometto} S.~P.,  2003, \apj, 588, 35

\bibitem[\protect\citeauthoryear{{Mandelbaum}, {Seljak}, {Kauffmann}, {Hirata}
  \& {Brinkmann}}{{Mandelbaum} et~al.}{2006}]{Man06}
{Mandelbaum} R.,  {Seljak} U.,  {Kauffmann} G.,  {Hirata} C.~M.,    {Brinkmann}
  J.,  2006, \mnras, 368, 715

\bibitem[\protect\citeauthoryear{{Navarro}, {Frenk} \& {White}}{{Navarro}
  et~al.}{1997}]{Nav97}
{Navarro} J.~F.,  {Frenk} C.~S.,    {White} S.~D.~M.,  1997, \apj, 490, 493

\bibitem[\protect\citeauthoryear{{Pierpaoli}, {Scott} \& {White}}{{Pierpaoli}
  et~al.}{2001}]{Pie01}
{Pierpaoli} E.,  {Scott} D.,    {White} M.,  2001, \mnras, 325, 77

\bibitem[\protect\citeauthoryear{{Prada}, {Klypin}, {Simonneau},
  {Betancort-Rijo}, {Patiri}, {Gottl{\"o}ber} \& {Sanchez-Conde}}{{Prada}
  et~al.}{2006}]{Pra06}
{Prada} F.,  {Klypin} A.~A.,  {Simonneau} E.,  {Betancort-Rijo} J.,  {Patiri}
  S.,  {Gottl{\"o}ber} S.,    {Sanchez-Conde} M.~A.,  2006, \apj, 645, 1001

\bibitem[\protect\citeauthoryear{{Prada}, {Vitvitska}, {Klypin}, {Holtzman},
  {Schlegel}, {Grebel}, {Rix}, {Brinkmann}, {McKay} \& {Csabai}}{{Prada}
  et~al.}{2003}]{Pra03}
{Prada} F.,  {Vitvitska} M.,  {Klypin} A.,  {Holtzman} J.~A.,  {Schlegel}
  D.~J.,  {Grebel} E.~K.,  {Rix} H.-W.,  {Brinkmann} J.,  {McKay} T.~A.,
  {Csabai} I.,  2003, \apj, 598, 260

\bibitem[\protect\citeauthoryear{{Press} \& {Schechter}}{{Press} \&
  {Schechter}}{1974}]{Pre74}
{Press} W.~H.,  {Schechter} P.,  1974, \apj, 187, 425

\bibitem[\protect\citeauthoryear{{Primack}}{{Primack}}{1997}]{Pri97}
{Primack} J.~R.,  1997, ArXiv Astrophysics e-prints (astro-ph/9707285)

\bibitem[\protect\citeauthoryear{{Primack}}{{Primack}}{2003}]{Pri03}
{Primack} J.~R.,  2003, Nuclear Physics B Proceedings Supplements, 124, 3

\bibitem[\protect\citeauthoryear{{Rines}, {Finn} \& {Vikhlinin}}{{Rines}
  et~al.}{2007}]{Rin07}
{Rines} K.,  {Finn} R.,    {Vikhlinin} A.,  2007, \apjl, 665, L9

\bibitem[\protect\citeauthoryear{{Sheldon}, {Johnston}, {Frieman}, {Scranton},
  {McKay}, {Connolly}, {Budav{\'a}ri}, {Zehavi}, {Bahcall}, {Brinkmann} \&
  {Fukugita}}{{Sheldon} et~al.}{2004}]{She04}
{Sheldon} E.~S.,  {Johnston} D.~E.,  {Frieman} J.~A.,  {Scranton} R.,  {McKay}
  T.~A.,  {Connolly} A.~J.,  {Budav{\'a}ri} T.,  {Zehavi} I.,  {Bahcall} N.~A.,
   {Brinkmann} J.,    {Fukugita} M.,  2004, \aj, 127, 2544

\bibitem[\protect\citeauthoryear{{Sheth}, {Mo} \& {Tormen}}{{Sheth}
  et~al.}{2001}]{She01}
{Sheth} R.~K.,  {Mo} H.~J.,    {Tormen} G.,  2001, \mnras, 323, 1

\bibitem[\protect\citeauthoryear{{Sheth} \& {Tormen}}{{Sheth} \&
  {Tormen}}{1999}]{She99}
{Sheth} R.~K.,  {Tormen} G.,  1999, \mnras, 308, 119

\bibitem[\protect\citeauthoryear{{Smith}, {Bernstein}, {Fischer} \&
  {Jarvis}}{{Smith} et~al.}{2001}]{Smi01}
{Smith} D.~R.,  {Bernstein} G.~M.,  {Fischer} P.,    {Jarvis} M.,  2001, \apj,
  551, 643

\bibitem[\protect\citeauthoryear{{Somerville} \& {Primack}}{{Somerville} \&
  {Primack}}{1999}]{Som99}
{Somerville} R.~S.,  {Primack} J.~R.,  1999, \mnras, 310, 1087

\bibitem[\protect\citeauthoryear{{Spergel}, {Bean}, {Dor{\'e}}, {Nolta},
  {Bennett}, {Dunkley}, {Hinshaw}, {Jarosik}, {Komatsu}, {Page}, {Peiris},
  {Verde}, {Halpern}, {Hill}, {Kogut}, {Limon}, {Meyer}, {Odegard}, {Tucker} \&
  {Weiland}}{{Spergel} et~al.}{2007}]{Spe07}
{Spergel} D.~N.,  {Bean} R.,  {Dor{\'e}} O.,  {Nolta} M.~R.,  {Bennett} C.~L.,
  {Dunkley} J.,  {Hinshaw} G.,  {Jarosik} N.,  {Komatsu} E.,  {Page} L.,
  {Peiris} H.~V.,  {Verde} L.,  {Halpern} M.,  {Hill} R.~S.,  {Kogut} A.,
  {Limon} M.,  {Meyer} S.~S.,  {Odegard} N.,  {Tucker} G.~S.,    {Weiland}
  J.~L.,  2007, \apjs, 170, 377

\bibitem[\protect\citeauthoryear{{Taylor} \& {Navarro}}{{Taylor} \&
  {Navarro}}{2001}]{Tay01}
{Taylor} J.~E.,  {Navarro} J.~F.,  2001, \apj, 563, 483

\bibitem[\protect\citeauthoryear{{The Dark Energy Survey Collaboration}}{{The
  Dark Energy Survey Collaboration}}{2005}]{Abb05}
{The Dark Energy Survey Collaboration} 2005, ArXiv Astrophysics e-prints (astro-ph/0510346)

\bibitem[\protect\citeauthoryear{{van Dokkum}, {Franx}, {Fabricant}, {Kelson}
  \& {Illingworth}}{{van Dokkum} et~al.}{1999}]{vDok99}
{van Dokkum} P.~G.,  {Franx} M.,  {Fabricant} D.,  {Kelson} D.~D.,
  {Illingworth} G.~D.,  1999, \apjl, 520, L95

\bibitem[\protect\citeauthoryear{{Wechsler}, {Bullock}, {Primack}, {Kravtsov}
  \& {Dekel}}{{Wechsler} et~al.}{2002}]{Wec02}
{Wechsler} R.~H.,  {Bullock} J.~S.,  {Primack} J.~R.,  {Kravtsov} A.~V.,
  {Dekel} A.,  2002, \apj, 568, 52

\bibitem[\protect\citeauthoryear{{White}}{{White}}{2001}]{Whi01}
{White} M.,  2001, \aap, 367, 27

\bibitem[\protect\citeauthoryear{{White} \& {Rees}}{{White} \&
  {Rees}}{1978}]{Whi78}
{White} S.~D.~M.,  {Rees} M.~J.,  1978, \mnras, 183, 341

\bibitem[\protect\citeauthoryear{{Wojtak}, {{\L}okas}, {Gottl{\"o}ber} \&
  {Mamon}}{{Wojtak} et~al.}{2005}]{Woj05}
{Wojtak} R.,  {{\L}okas} E.~L.,  {Gottl{\"o}ber} S.,    {Mamon} G.~A.,  2005,
  \mnras, 361, L1

\bibitem[\protect\citeauthoryear{{Wojtak}, {Lokas}, {Mamon}, {Gottloeber},
  {Klypin} \& {Hoffman}}{{Wojtak} et~al.}{2008}]{Woj08}
{Wojtak} R.,  {Lokas} E.~L.,  {Mamon} G.~A.,  {Gottloeber} S.,  {Klypin} A.,
  {Hoffman} Y.,  2008, ArXiv e-prints (arXiv:0802.0429)

\bibitem[\protect\citeauthoryear{{Zaritsky} \& {White}}{{Zaritsky} \&
  {White}}{1994}]{Zar94}
{Zaritsky} D.,  {White} S.~D.~M.,  1994, \apj, 435, 599

\bibitem[\protect\citeauthoryear{{Zhao}, {Mo}, {Jing} \& {B{\"o}rner}}{{Zhao}
  et~al.}{2003}]{Zha03}
{Zhao} D.~H.,  {Mo} H.~J.,  {Jing} Y.~P.,    {B{\"o}rner} G.,  2003, \mnras,
  339, 12

\end{thebibliography}

\bsp
\end{document}